\documentclass[]{aastex7}
\usepackage{hyperref}
\usepackage{amsmath}
\usepackage{amssymb}
\usepackage{graphicx}
\usepackage{gensymb}
\usepackage{cleveref}
\usepackage{pifont}

\usepackage{footnote}
\usepackage{placeins}
\usepackage{float}

\graphicspath{{./}{figures/}}

\newcommand{\urlfootnote}[1]{\footnote{\raggedright\url{#1}}}

\submitjournal{Publications of the Astronomical Society of the Pacific}

\begin{document}

\title{GReX: An Instrument Overview and New Upper Limits on the Galactic FRB Population}

\correspondingauthor{Kiran A. Shila}

\author[0000-0003-4652-7038]{Kiran A. Shila}
\affiliation{Cahill Center for Astronomy and Astrophysics, California Institute of Technology, Pasadena, CA 91125, USA}
\email[show]{kshila@caltech.edu}

\author[0009-0006-5007-7470]{Sashabaw Niedbalski}
\affiliation{Department of Astronomy and Center for Astrophysics and Planetary Science, Cornell University, Ithaca, NY 14853, USA}
\email{sn526@cornell.edu}

\author[0000-0002-7587-6352]{Liam Connor}
\affiliation{Center for Astrophysics Harvard \& Smithsonian, 60 Garden Street, Cambridge, MA 02138, USA}
\email{liam.connor@cfa.harvard.edu}

\author[0000-0001-5390-8563]{Shrinivas R. Kulkarni}
\affiliation{Cahill Center for Astronomy and Astrophysics, California Institute of Technology, Pasadena, CA 91125, USA}
\email{srk@astro.caltech.edu}

\author[0009-0007-4568-3115]{Larom Segev}
\affiliation{Center for Astrophysics Harvard \& Smithsonian, 60 Garden Street, Cambridge, MA 02138, USA}
\email{laromsegev@college.harvard.edu}

\author[0009-0001-2976-6486]{Priya Shukla}
\affiliation{Cahill Center for Astronomy and Astrophysics, California Institute of Technology, Pasadena, CA 91125, USA}
\affiliation{Raman Research Institute, Bangalore, India}
\email{shuklapriya244@gmail.com}

\author[0000-0002-4553-655X]{Evan F. Keane}
\affiliation{School of Physics, Trinity College Dublin, College Green, Dublin 2, Ireland}
\email{evan.keane@tcd.ie}

\author[0000-0003-4399-2233]{Joseph McCauley}
\affiliation{School of Physics, Trinity College Dublin, College Green, Dublin 2, Ireland}
\email{joe.mccauley@tcd.ie}

\author[0000-0002-5927-0481]{Owen A. Johnson}
\affiliation{School of Physics, Trinity College Dublin, College Green, Dublin 2, Ireland}
\affiliation{Radio Astronomy Laboratory, University of California, Berkeley, CA, 94720 USA}
\email{ojohnson@tcd.ie}

\author[0009-0006-3224-4167]{Brendan Watters}
\affiliation{School of Physics, Trinity College Dublin, College Green, Dublin 2, Ireland}
\email{wattersb@tcd.ie}

\author[0000-0002-0161-7243]{Wael Farah}
\affiliation{SETI Institute, 339 Bernardo Ave, Suite 200 Mountain View, CA 94043, USA}
\email{wael.a.farah@gmail.com}

\author[0000-0002-3430-7671]{Alexander W. Pollak}
\affiliation{SETI Institute, 339 Bernardo Ave, Suite 200 Mountain View, CA 94043, USA}
\email{alexander.pollak.87@gmail.com}

\author{Konstantin Belov}
\affiliation{Cahill Center for Astronomy and Astrophysics, California Institute of Technology, Pasadena, CA 91125, USA}
\email{konstantin.v.belov@jpl.nasa.gov}

\author[0009-0003-9453-2779]{Honglue Tang}
\affiliation{Center for Astrophysics Harvard \& Smithsonian, 60 Garden Street, Cambridge, MA 02138, USA}
\email{calnatns@whu.edu.cn}

\author[0009-0009-1219-5128]{Zhaoyu Huai}
\affiliation{Cahill Center for Astronomy and Astrophysics, California Institute of Technology, Pasadena, CA 91125, USA}
\email{zhuai@caltech.edu}

\author[0000-0002-2878-1502]{Shami Chatterjee}
\affiliation{Department of Astronomy and Center for Astrophysics and Planetary Science, Cornell University, Ithaca, NY 14853, USA}
\email{shami.chatterjee@cornell.edu}

\author[0000-0002-4049-1882]{James M. Cordes}
\affiliation{Department of Astronomy and Center for Astrophysics and Planetary Science, Cornell University, Ithaca, NY 14853, USA}
\email{jmc33@cornell.edu}

\begin{abstract}
We present the instrument design and initial results for the Galactic Radio Explorer (GReX), an all-sky monitor for exceptionally bright transients in the radio sky.
This instrument builds on the success of STARE2 to search for fast radio bursts (FRBs) from the Milky Way and its satellites.
This instrument has deployments across the globe, with wide sky coverage and searching down to $32\,\mu\text{s}$ time resolution, enabling the discovery of new super giant pulses.
Presented here are the details of the hardware and software design of the instrument, performance in sensitivity and other key metrics, and experience in building a global-scale, low-cost experiment.
We follow this discussion with experimental results on validation of the sensitivity via hydrogen-line measurements.
We then update the rate of Galactic FRBs based on non-detection in the time since FRB 200428. 
Our results suggest FRB-like events are even rarer than initially implied by the detection of a MJy burst from SGR J1935+2154 in April 2020.
\end{abstract}


\section{Introduction} \label{sec:intro}

Fast radio bursts (FRBs) are short-duration ($\lesssim 100$\,ms), highly energetic ($\gtrsim 10^{40}$\,erg\,s$^{-1}$) transients that have been detected between 100\,MHz and 8\,GHz \citep{cordes-review, petroffreview}. 
Prior to April 2020, all FRBs were extragalactic and most were at cosmological distances ($z\geq0.1$).
The discovery of SGR\,1935+2154 by both STARE2 and the Canadian Hydrogen Intensity Mapping Experiment (CHIME/FRB) made a dramatic connection between extragalactic FRBs and Galactic magnetars \citep{andersen_bright_2020,bochenekSTARE2DetectingFast2020}. 
While FRB\,200428 from SGR\,1935+2154 was less luminous than all known extragalactic FRBs, its proximity made it the highest fluence of any exosolar radio burst ($\sim$\,MJy\,ms at 1\,GHz, compared to $\lesssim$\,kJy\,ms at $\sim$1\,GHz for high fluence extragalactic FRBs \citep{kirsten_link_2024}) and enabled detection of the first prompt multi-wavelength emission \citep{integralFRB}.

In the years since the FRB-like emission from SGR\,1935+2154, the link between young, energetic magnetars and the extragalactic FRB phenomenon has been muddied.
FRBs have been found in globular clusters \citep{kirsten_repeating_2022}, at the outskirts of elliptical galaxies \citep{chime_elliptical_1, chime_elliptical_2}, and may be underrepresented in low-mass star forming galaxies and the sites of recent core-collapse supernovae \citep{sharma24}. In other words, if all FRBs are magnetars, they are not all young remnants of CCSNe embedded in star-forming region, like SGR\,1935+2154.
These facts motivate a blind search survey, rather than monitoring known Galactic magnetars.
The discovery of other new Galactic phenomena such as long-period radio transients \citep{HurleyWalker, Caleb, RodriguezLPRT} further motivates blind searches in new regions of parameter space. 

The Galactic Radio Explorer is an international collaboration aimed at discovering exceptionally bright bursts in the radio sky.
Building off the success of STARE2~\citep{bochenekSTARE2DetectingFast2020}, we are currently searching for FRBs emitted by Galactic sources as well as nearby galaxies.
We search down to a high time resolution of $32.768\,\mu\text{s}$, saving raw $8.192\,\mu\text{s}$ voltage data to enable more in-depth analysis for interesting single-pulse candidates.
Additionally, the instrument has been designed to be easy to build and reproduce, allowing collaborators to quickly bring up a functioning station.
The goal as mentioned in our original white paper~\citep{connorGalacticRadioExplorer2021a} is to eventually have 4$\pi$ steradian coverage with increased exposure to the galactic plane to improve sensitivity to galactic sources.
As more sensitive aperture arrays for fast transient discovery come online, such as BURSTT \citep{BURSTT} in Taiwan and similar efforts in the US, Australia, and Chile, the niche of GReX will be all-sky 
sensitivity to ultra-narrow, bright radio bursts at 1.4\,GHz.
Each GReX terminal is fully self-contained and low-cost, serving as a valuable pedagogical tool for the observatory or university operating it locally.

In this work, we describe the GReX instrument, and the current network of terminals around the world. We also place new upper-limits on the rate of ultra-bright FRBs based on non-detection.
\section{Hardware Design} \label{sec:hardware}

\subsection{Analog Signal Path}
The first major hardware improvement over STARE2 is the transition to a new, higher performance low noise amplifier (LNA)~\citep{weinrebLowNoiseAmplifier2021a} developed at Caltech for the DSA-110~\citep{raviDSA110OverviewFirst2023}.
The new amplifier has a noise temperature of $7\,\text{K}$, versus the previous amplifier's noise temperature of $\approx30\,\text{K}$, dramatically improving our system noise temperature and sensitivity.
The LNA achieves record-breaking noise at ambient temperature via the use of an exceptional low noise transistor from Diramics and a low loss suspended stripline input matching network.

For the antenna, we use an improved version of the same ``cake pan'' antenna from STARE2, also designed for the DSA-110.
This is a waveguide horn antenna with cake pans forming axial corrugations, enhancing the beam pattern.
The antenna's full-width half-maximum (FWHM) is estimated to be $70\degree\pm5\degree$ at $1.4\,\text{GHz}$ with very little variation across our band of 1280 to 1530 MHz.
We verify the estimation of the beam width in \autoref{sec:beam_model} by fitting HI intensity data.
As this is an all-sky instrument, the large beam width trades sensitivity for field of view, matching our science goals.

\subsection{Frontend Module}
\begin{figure}[ht!]%
    \centering
    \parbox{.25\textwidth}{
        \centering
        \includegraphics[width=\linewidth]{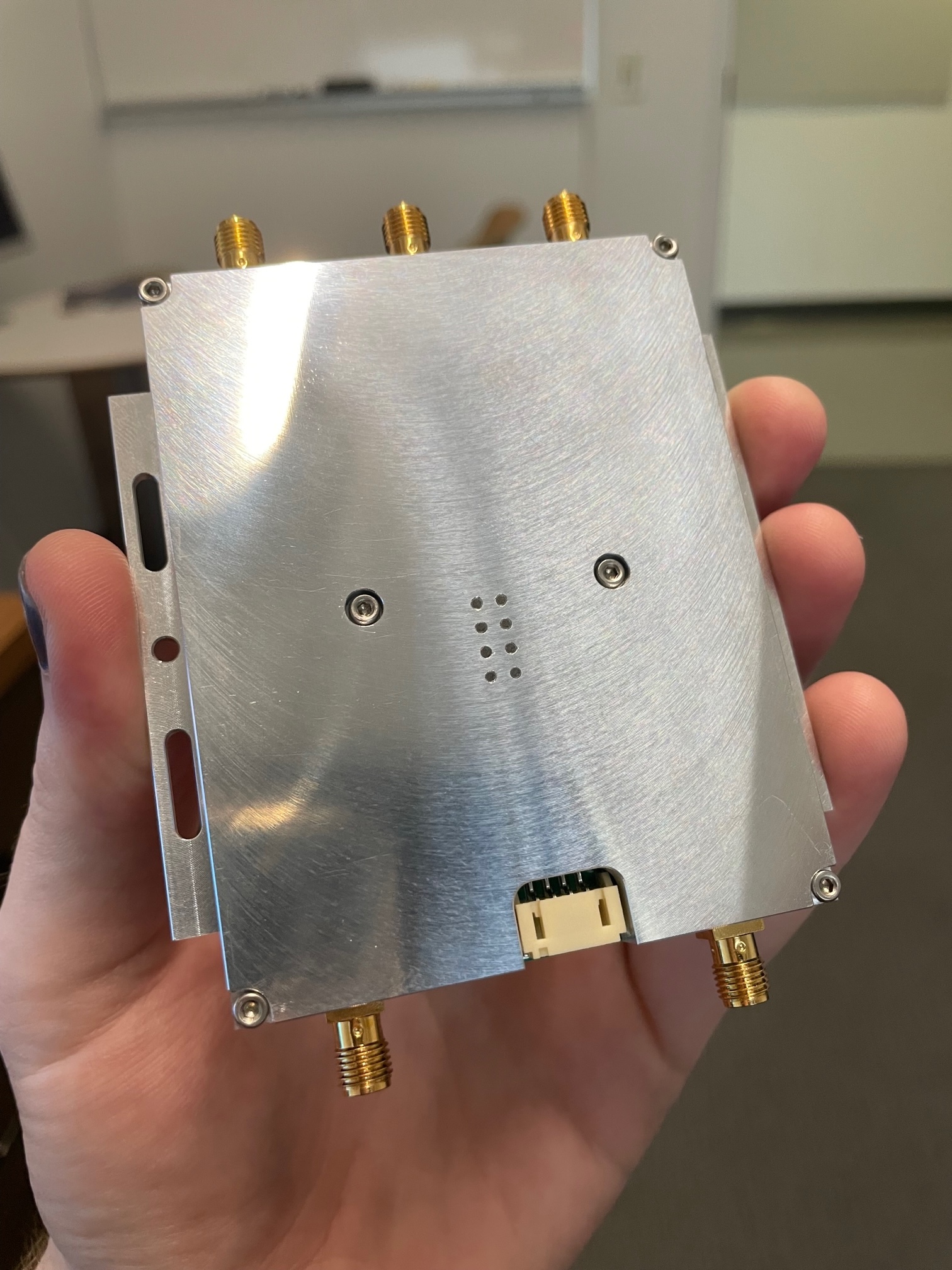}
    }%
    \qquad
    \parbox{.25\textwidth}{
        \centering
        \includegraphics[width=\linewidth]{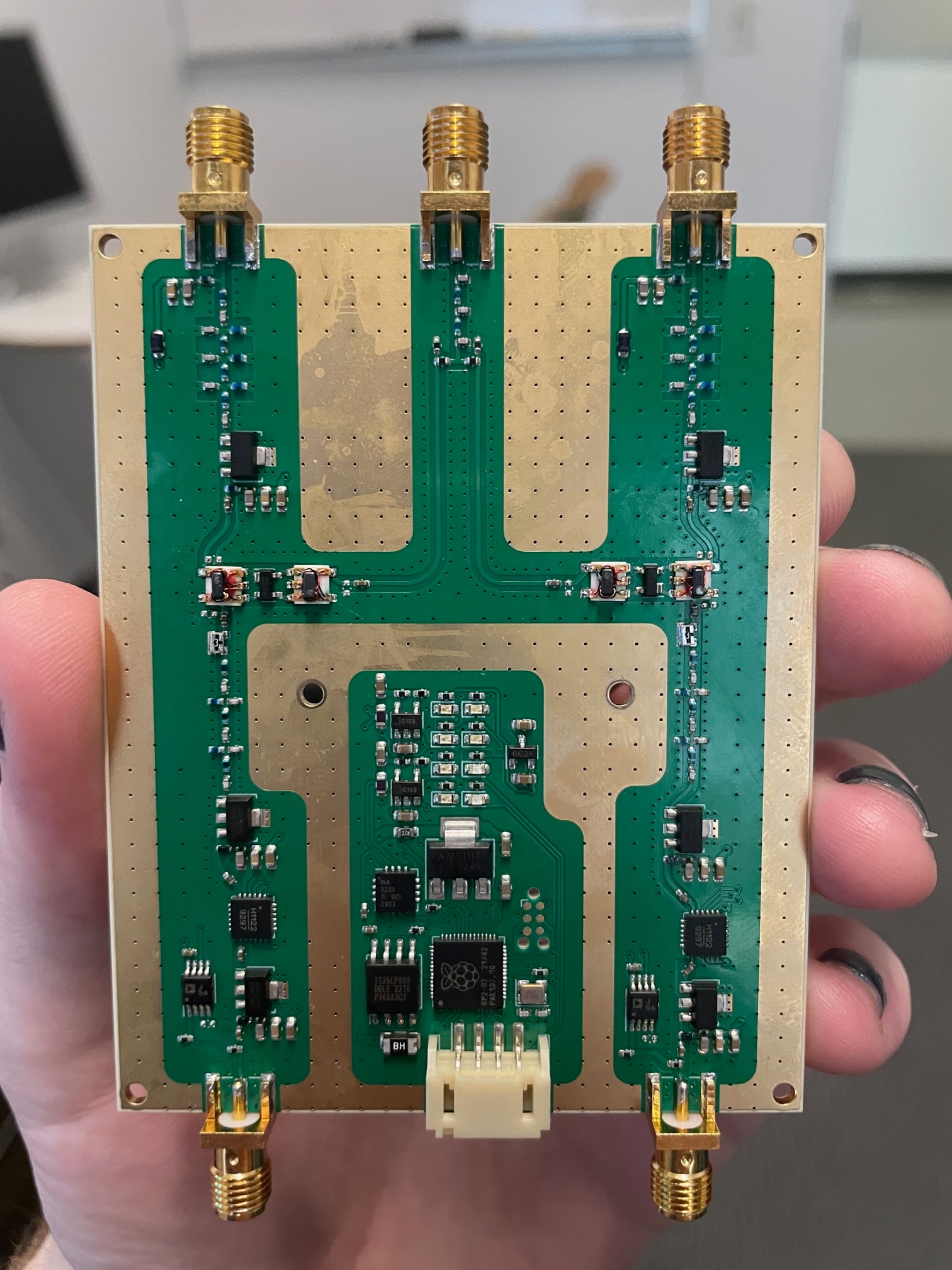}
    }%
    \caption{Completed FEMs}%
    \label{fig:fem}%
\end{figure}

The heart of the analog signal processing in GReX is the frontend module (FEM).
In STARE2, the analog processing was split between electronics at the antenna and in the server room, with an RF over fiber (RFoF) link between the two sides.
Instead of RFoF, we digitize directly at the antenna, incorporating all the electronics into a single enclosure to which the antenna is bolted.
As RFoF links are typically quite nonlinear, this section of the analog signal path could limit the linear dynamic range.
To replace the RFoF circuitry, we designed the FEM to perform the frequency downconversion, filtering, and amplification needed to prepare the signal for digitization.

The FEM also contains digital control electronics using an RP2040 microcontroller.
This device controls the power for the LNAs connected to the FEM inputs, sets the variable attenuators, and monitors temperatures and average RF power levels.
We wrote the firmware in the Rust programming language using the RTIC~\citep{erikssonRealtimeMassesStep2013} framework for real-time task-based concurrency.
Monitoring and control is performed via a simple serial connection.

\begin{figure}[t!]
    \centering
    \includegraphics[width=0.7\linewidth]{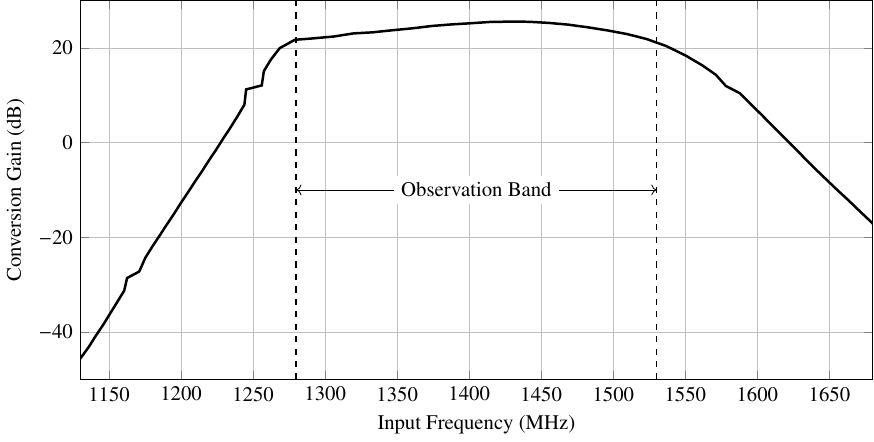}
    \caption[Measured conversion gain of the FEM]{Measured conversion gain of the FEM. Total system gain adds $40\,\text{dB}$ from the LNA and $15\,\text{dB}$ from ADC preamplifiers}
    \label{fig:fem_bandpass}
\end{figure}

The FEM enclosure is custom machined with channels that isolate the two polarizations from the digital electronics.
This enclosure also acts as the heat sink, as it is bolted directly to a large aluminum panel inside the box.
The FEMs were fully factory-assembled, with the total per-unit cost under \$200 USD.
The completed FEM enclosure and PCB are shown in \autoref{fig:fem}.
The conversion gain of the FEM is shown in \autoref{fig:fem_bandpass}.

\subsection{The Box}

\begin{figure}[ht!]
    \centering
    \includegraphics[width=0.6\linewidth]{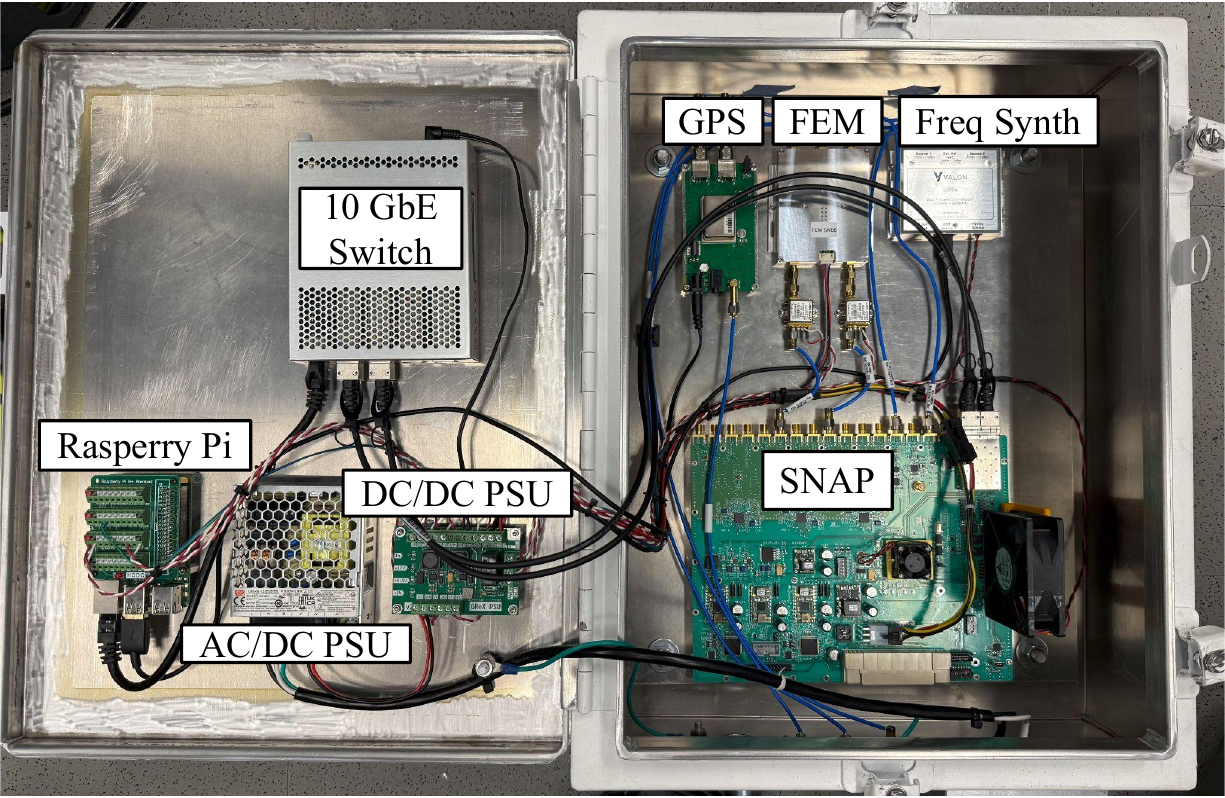}
    \caption[Inside a GReX box]{Inside a GReX box. The left panel contains the 10 GbE switch, two power supply units (PSU), and the Raspberry Pi for monitor and control. The right half contains the GPS timing receiver, frontend module (FEM), frequency synthesizer, and SNAP digitizer.}
    \label{fig:box_inside}
\end{figure}

The constructed telescope is contained in a single weatherproof aluminum box $20^{\prime\prime}$ tall, $16^{\prime\prime}$ wide, and $6^{\prime\prime}$ deep. 
This box contains a GPS timing system and oscillators, the FEM, the FPGA digitizer, power supplies, a 10G Ethernet switch, and a Raspberry Pi for monitor and control.
As the box we chose had no inherent RFI shielding, we added an aftermarket weatherproof RFI gasket.
The FEM, synthesizer, GPS receiver, and FPGA digitizer are bolted to a subpanel inside the box.
The two polarization connections and GPS antenna are connected via feedthrough connectors on the bottom of the box.
The optical fiber is routed through a waveguide cutoff pipe that is attached to the bottom of the box, where it is internally terminated to a 10G SFP+ connector, attached to the Ethernet switch.
The box has mounting points on the outside for a stand or pipe and boltholes to mount the antenna.
The antenna is bolted on the top of the box with weatherproofing gasket material to prevent any water leaks.

Installation of the box is indented to be simple, requiring two single-mode fibers for the Ethernet connection and mains AC power.
Upon receiving a box, the operator needs to bolt the antenna to the top with the weatherproofing compound, then mount it to an appropriate place with a clear view of the sky.
Ideally, the antenna should be covered with RF-transparent material for weatherproofing, but we found any piece of low-loss material such as a plastic trash can to be sufficient.
A picture of an assembled box is shown in \autoref{fig:box_inside}.
\section{Software Design} \label{sec:software}

As this experiment is deployed globally, it is imperative that the software stack be easy to build, install, and use, all with very little assistance from the project maintainers.
New members should be able to on-board their stations independently, just by following online documentation\urlfootnote{https://grex-telescope.github.io}.
As such, we developed the software stack for this instrument with intense attention to these goals.
As much as possible, software is tested in continuous integration and written in a way to reduce the likelihood of unexpected errors.
Additionally, all the software for the project is freely accessible and under an open-source license.

\subsection{Digitization and Initial Processing}

Following the analog signal path, the high-frequency signals are digitized by the Smart Network ADC Processor (SNAP)\urlfootnote{https://casper.berkeley.edu/wiki/SNAP} board.
This platform contains three HMCAD1511 analog to digital converters (ADC) from Analog Devices and a Kintex-7 160T field-programmable gate array (FPGA) from Xilinx.
This platform is supported by the CASPER ~\citep{hickishDecadeDevelopingRadioAstronomy2016a} ecosystem, which we make use of here.

We wrote our gateware\urlfootnote{https://github.com/GReX-Telescope/gateware} in a combination of Simulink and SystemVerilog using the CASPER toolflow.
To make maximum use of the available 10G connection, we stream 8+8 bit complex data for each polarization, channelized to 2048 frequency bins.
As the FPGA core is clocked at 250 MHz, with the ADC clock running 500 MHz, the F-engine processes two ADC samples every clock cycle.
The channelization is accomplished with a standard polyphase filterbank constructed via the combination of a finite impulse response (FIR) filter and fast Fourier transform (FFT), both making use of design blocks from the CASPER library.
The FIR filter has 8 taps and a Hamming windowing function to reduce inter-channel leakage and scalloping loss~\citep{Thompson2017}.
To avoid overflows in the FFT, each stage is allowed to grow the number of bits representing each channel.
The output of the FFT block is 18+18 bit, fixed-point complex numbers.
If we attempted to stream these data directly, we would not have the bandwidth on the 10G link.
As such, these data are requantized to a lower bit-depth of 8+8 bits to fit in a single 10G Ethernet payload.
Finally, the resulting channelized voltage data is packetized for transmission.
The block of data includes a 64-bit packet counter header, allowing the downstream software to detect data loss and re-ordering.
These data were packed such that they match the layout of a C-struct, so no processing needs to be done to use the data downstream.
The packetized data is transmitted over the optical 10G Ethernet link to the server, making use of the standard Ethernet block from CASPER.
Every 2048 samples, a completed Ethernet payload is transmitted, including the header, totalling 8200 bytes.
This occurs every 8.192 us, implying a total data rate of just over 1 GB/s or 8 GiB/s.

\begin{figure}[t!]
    \centering
    \includegraphics[width=0.7\linewidth]{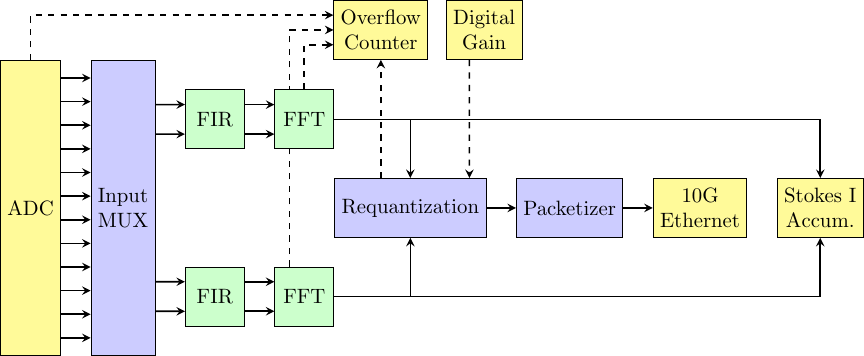}
    \caption{FPGA gateware internals: yellow blocks represent CASPER-provided interfaces to hardware, blue blocks are custom SystemVerilog, and green blocks are CASPER-provided DSP logic. Solid lines represent high-speed data, dashed lines represent slow monitor and control data. Reset lines and timing logic is not shown.}
    \label{fig:gateware}
\end{figure}

In addition to the primary functionality of the gateware, there are a few added utilities to assist in operation.
First, there is an input multiplexer that allows the user to connect to any input on the SNAP without reprogramming.
Second, there is a triggerable accumulator for Stokes I data that gives the user a snapshot view of the spectra without running high-speed packet capture.
Additionally, this accumulation is performed on the raw output from the FFT block, providing insight into the required digital gain setting for requantization.
Finally, there are various registers that monitor overflow conditions in both requatization, FFT bit growth, and raw ADC samples.
This primarily gives us an indication that the total power level into the SNAP board is too high, or downstream digital gain is too high, either case being actionable via the adjustable RF attentuators in the FEM or in the digital gain settings.
A figure of the data flow in the gateware is shown in \autoref{fig:gateware}.

\subsection{FPGA Software Interface}

A critical piece of code in the project is the software interface to the running FPGA gateware.
During telescope operation, status registers and accumulators are read, variables are set, etc.
For this project, we chose to write a new interface to CASPER devices using the Rust programming language, \textit{casperfpga\_rs}\urlfootnote{https://github.com/kiranshila/casperfpga_rs}.
The new library generates compile-time checked interfaces to the various components in the running FPGA design.
As the generated code fully validates correctness at compile-time, runtime errors are all but eliminated.

This new library allows for much higher confidence in the deployment of CASPER designs.
This library also contains unit tests against a mock interface, extensive documentation, and is written in a modular nature to encourage other members of the CASPER collaboration to add support for their hardware.

\subsection{Server and First Stage Processing} \label{subsec:t0}

After packets have been transmitted from the FPGA, they are captured and processed by a single high-performance server.
This server contains a 24-core AMD Ryzen Threadripper Pro 5965WX with 128 GB of RAM.
The server is also outfitted with an NVIDIA GeForce RTX 3090 Ti GPU, used for the brute-force dedispersion search.
This system's kernel parameters are tuned to support the “jumbo Ethernet frames” emitted from the SNAP board.
After the packets have been processed through the kernel, user-space programs perform the remaining processing.

The majority of the effort in software development for this project was in the first stage processing program, \textit{T0}.
This program performs the packet capture, computation of Stokes I, time downsampling, and management of ring buffers of voltage data to be written to disk on command.
As this piece of software handles the raw voltage data, it is the most timing sensitive and performance-critical.

At the top level, \textit{T0} consists of many independent threads, each performing a processing step called a subtask.
One thread is dedicated to the packet capture subtask, one for time-downsampling, etc.
We must use multithreading as the total processing time is longer than the incoming packet cadence, so parts of the processing need to be broken up and performed in parallel, i.e., pipelined.
After each subtask completes computation, it passes the result to the subsequent subtask across a thread boundary.

Our implementation of this streaming processing pipeline is built off the inter-process communication model of channels with allocation-reusing ring buffers, similar in spirit to PSRDADA~\citep{vanstratenPSRDADADistributedAcquisition2021}.
Unlike PSRDADA, the model is fully contained in a single executable, allowing for much stronger guarantees around the data being passed around.
Additionally, this model reduces the need for synchronization primitives such as mutexes, improving performance. 
\begin{figure}[t]
    \centering
    \includegraphics[width=0.7\linewidth]{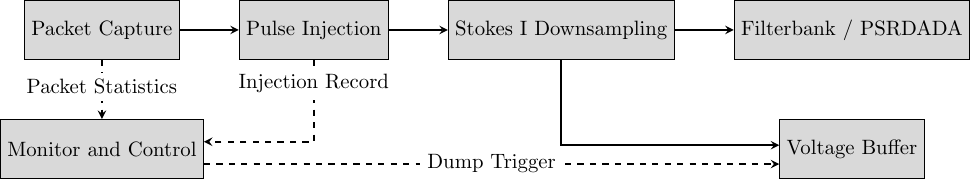}
    \caption{\textit{T0} subtasks and inter-thread communication. Solid lines represent high-speed science data, dashed lines represent monitor and control.}
    \label{fig:t0_subtasks}
\end{figure}

The complete architecture of \textit{T0} is shown in \autoref{fig:t0_subtasks}, with arrows showing the channels that move data across the various threads.
After program start and timing synchronization (described in the \autoref{sec:timing}), all the subtasks are spawned simultaneously.
The first subtask again is packet capture, where incoming data is read from the network card.
This subtask checks the value of the packet header against the previous to test for packet loss or reordering.
Metadata about the packet statistics is transferred using a separate channel to the monitoring thread.
The following subtask performs fake pulse injection.
As we want to ensure our pipeline is working properly, we occasionally add synthetic data of various dispersion measures and fluence into the real-time data stream.
We write information about the injection into a \textit{SQLite} database, so downstream processing software can determine if a given candidate is synthetic.
The next subtask performs time downsampling.
Downstream processing, specifically brute-force dedispersion, has trouble handling exceptionally high time resolution data.
As such, we need to perform downsampling in time, in addition to computing Stokes I as we will be performing incoherent dedispersion.
Finally, depending on launch arguments, the downsampled Stokes I data is written to either a \textit{SIGPROC}\urlfootnote{https://sigproc.sourceforge.net/} filterbank file (using a high-performance Rust implementation\urlfootnote{https://github.com/kiranshila/sigproc_filterbank} of the file format) or to a \textit{PSRDADA}\urlfootnote{https://psrdada.sourceforge.net/} ring buffer.

Running in parallel to the data processing pipeline, a thread is accumulating and distributing monitor information.
This data includes occasional queries to the FPGA about overflow statistics, long-integration Stokes I spectra, and internal temperatures.
Moreover, it collects log messages produced by the program and statistics about packet capture from the first subtask.
The monitor subtask provides an HTTP API to query the monitor information for a Prometheus\urlfootnote{https://prometheus.io/} time-series database.
Log messages and traces are ingested by an \textit{OpenTelemetry}\urlfootnote{https://opentelemetry.io/} collector.

\subsection{Data Loss}
As the packets are transmitted over Ethernet with the UDP protocol, there is no guarantee that they are received.
We have taken special care to ensure the server's operating system can efficiently process incoming UDP data without data loss and \textit{T0} should be able to process incoming data in real time without issue.
However, packet loss still does occur, albeit rarely.
For the Owens Valley station, the average packet loss rate is $10^{-6}$\% or one packet dropped per $10^8$ packets processed.
As our packet cadence is one packet per $8.192\,\mu\text{s}$, this works out to an average of one dropped packet per $1000\,\text{s}$.
This amount of data loss is to be expected and is inconsequential to the performance of the telescope.

\subsection{Timing and Synchronization} \label{sec:timing}
In the box, we include a GPS-discriminated $10\,\text{MHz}$ oscillator.
This device provides a stable reference and emits a pulse-per-second (PPS) square wave, connected to a general purpose digital input on the FPGA.
The $10\,\text{MHz}$ reference clock is used by a Valon 5009a frequency synthesizer to produce the $500\,\text{MHz}$ clock used by the ADC and FPGA.
Furthermore, this synthesizer is used to generate the $1030\,\text{MHz}$ local oscillator signal used in the FEM for downconversion.

In the FPGA gateware, the PPS signal is used to timestamp data.
In \textit{T0} at program start, we query a network timeserver to get the current local time within tens of milliseconds.
Then, the program waits until the next half second to ``arm'' the timing system on the FPGA.
Once the next rising edge of the PPS signal arrives in the FPGA and starts the flow of data, we know that the very first voltage sample is coincident with the next whole second.

As each packet has a 64 bit packet counter, and we know the timestamp of the first packet and know the clock speed of the FPGA, we can work out the start time of every packet.
Specifically, we know that packet $N$ is exactly $N\cdot8.192\,\mu\text{s}$ past the time of the first packet.
This information is used in the metadata stored in voltage dumps as well as in the candidate metadata.

\subsection{RFI Cleaning} \label{subsec:rfi_clean}

Following \textit{T0}, data is transferred to an RFI cleaning program, \textit{clean\_rfi}\urlfootnote{https://github.com/GReX-Telescope/clean_rfi} using \textit{PSRDADA} ring buffers.
This program implements a multistep cleaning approach following the implementation in CHIME/FRB~\citep{rafiei-ravandiMitigatingRadioFrequency2023}.

For both \textit{T0} and \textit{clean\_rfi}, we needed a Rust interface to the \textit{PSRDADA} C library, as that is the only mechanism to stream data into the brute-force dedispersion program, \textit{HEIMDALL}~\citep{barsdellAcceleratingIncoherentDedispersion2012}.
While Rust has a robust interface to C programs and libraries, they are all memory-unsafe by default, as there is no mechanism to guarantee Rust's invariants across the application binary interface (ABI) boundary.
As such, we wrote a high-level Rust wrapper library called \textit{psrdada\_rs}\urlfootnote{https://github.com/kiranshila/psrdada-rs}.
Much like \textit{casperfpga\_rs}, this library adds a significant number of compile-time checks to guarantee correct usage before runtime.

\textit{clean\_rfi} implements simple masking, variance-cut, and detrending algorithms to iteratively remove RFI from a block of dynamic spectra.
Specifically, we start with a static mask of bad channels for frequencies that are consistently contaminated.
Then, we remove all samples that contain zeros, as the typical noise floor will be just above zero and pure-zeros represent dropped packets.
Next, we remove bandpass variation by dividing each sample by the time-average frequency response.
Finally, we perform variance cuts in both the frequency and time axes by removing data above some $\sigma$ threshold.
We perform these final variance cuts twice, with a higher $\sigma$ in the second pass.
Currently, the iterative variance cut process removes time samples / frequency channels that are greater than $3\sigma$ and then $5\sigma$.

\subsection{Real-Time Detection Pipeline}\label{sec:realtime}

\begin{figure}[t!]
    \centering
    \parbox[t][][t]{.25\textwidth}{
        \centering
        \includegraphics[height=6cm]{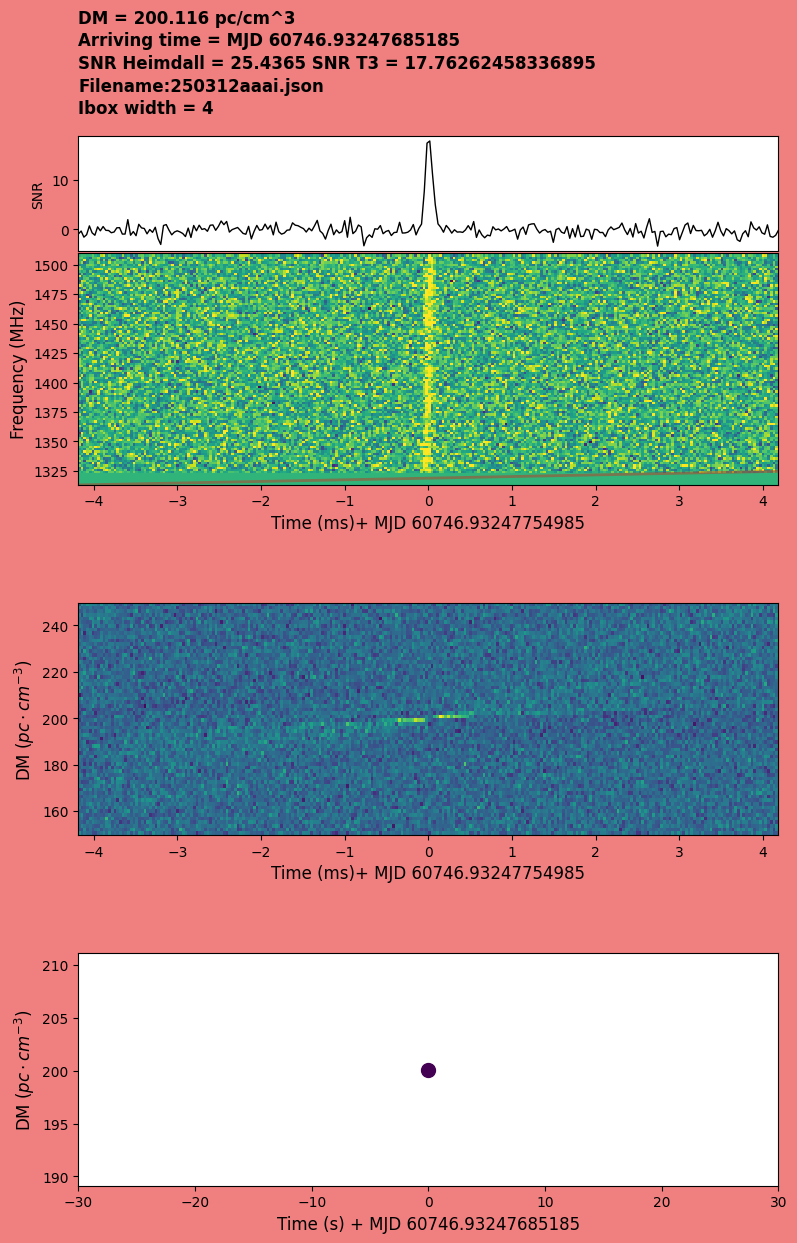}
        \caption{Detection candidate for an injected burst}%
        \label{fig:injection_cand}
    }\quad
    \parbox[t][][t]{.65\textwidth}{
        \centering
        \includegraphics[height=6cm]{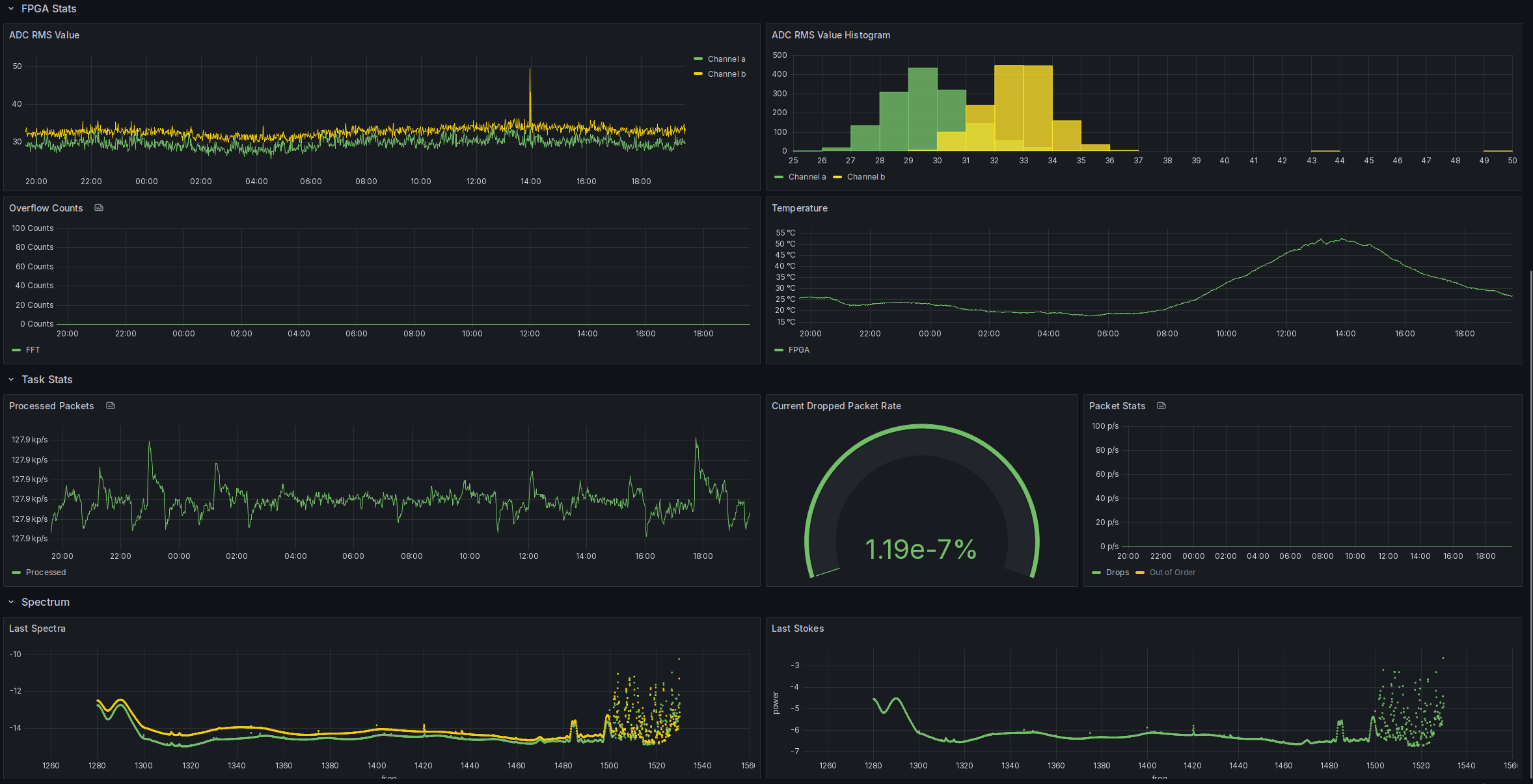}
        \caption{Grafana dashboard monitoring system performance}%
        \label{fig:ovro-dash}
    }
\end{figure}

After the dynamic spectra data were cleaned, they are passed along to our fork\urlfootnote{https://github.com/GReX-Telescope/heimdall-astro} of HEIMDALL.
We use HEIMDALL, and the associated \textit{dedisp} library, as the implementation of the brute-force dedispersion search.
Our fork removes the built-in RFI cleaning routines, removes candidate clustering, adds a structured logging library, and enables writing candidates to a network socket instead of a file.
As we will be running the search in real time, the included RFI cleaning and clustering routines were too slow for our high time-resolution data.

HEIMDALL then writes lines of candidates over a local network socket to the candidate filtering task, \textit{T2}.
This program is a fork of the \textit{T2} Python project written for DSA-110, modified to work with our data formats.
This program ingests the stream of candidates emitted by HEIMDALL, clusters them using HDBSCAN~\citep{McInnes2017}, filters the clustered results in dispersion measure, SNR, and boxcar width to produce candidates.
Additionally, for every viable candidate, a message is sent to \textit{T0} to dump the contents of the voltage ring buffer to disk.

\textit{T3} is the final stage in the pipeline, another fork from the DSA-110 project.
This program watches the directory where candidate files are written and generates plots of the dedispersed, RFI-cleaned dynamic spectra.
These candidate plots are then pushed to a Slack channel, where members of the collaboration get real-time notifications.
An example of this notification is shown in \autoref{fig:injection_cand}.
Eventually, \textit{T3} will perform machine learning-based candidate classification as well as inter-station communication for coincidencing and localization.

The whole pipeline is orchestrated through a single bash script.
This script launches all the tasks sequentially using GNU Parallel~\citep{tange_2025_14911163}, including the initial setup of the PSRDADA buffers.
The script has several launch modes, allowing the operator to run the full processing pipeline, stream into a named PSRDADA buffer, stream into a filterbank file, etc.
Once the detection pipeline has sent a trigger, \textit{T0} writes a self-describing NetCDF binary file of voltage data to disk where the candidate can be processed offline.

\subsection{Technosignature Searches}

GReX units can also be used to carry out technosignature surveys.
GReX's large field of view enables the constraining of upper limits on the prevalence of intelligent life in the local galaxy.
Presently, two technosignature pipelines can be deployed. 
A high-spectral-resolution technosignature detection pipeline can be deployed to search for drifting narrowband technosignatures.
In the case of GReX, coincidence rejection from two or more units can be used to mitigate non-terrestrial narrowband signals \citep{johnson_simultaneous_2023}.
Additionally, artificially dispersed signals can be searched for, as described in \cite{gajjar_searching_2022}, by deploying \textsc{SPANDAK}\footnote{\url{https://github.com/gajjarv/PulsarSearch/}} on GReX.
The minimum power of a transmitter that can be detected is given in terms of effective isotropic radiated power (EIRP), for a narrowband signal, this is expressed as

\begin{equation}
    \text{EIRP}_\text{min,narrow}(f, l, b) = \sigma \cdot 4\pi d_\star^2  \frac{2 k_b T_{\text{sys}}(l, b)}{A_e(f)}  \frac{1}{\sqrt{n_p  t_\text{obs}  \delta\nu}}.
\end{equation}

Here, $\sigma$ is the required signal-to-noise ratio (SNR), $\delta\nu$ is the bandwidth of the received signal, $t_{\text{obs}}$ is the observing integration time, $A_e(f)$ effective area of the telescope as function of frequency, $T_{sys}(l,b)$ system temperature as a function of sky position, $n_p$ is the number of polarizations, and $d_\star$ is the distance between the transmitter and receiver, i.e., the distance to the star.
A value of 1~Hz is assumed for $\delta\nu_t$. For narrowband signals considered in Doppler searches. Similarly, an artificially dispersed signal can be expressed as
\begin{equation}
    \text{EIRP}_\text{min,disp}(f, l, b) = \sigma \cdot 4\pi d_\star^2  \dfrac{2 k_b T_{\text{sys}}(l, b)}{A_e(f)}  \dfrac{1}{\sqrt{n_p ~\delta \nu ~ \tau}}.
\end{equation}

However, in this case $\tau$ represents the pulse duration of the dispersed burst where 1~ms is assumed.
A technosignature emanated from Alpha Centauri (1.34~pc) would require $1.2\times10^{12}\,\text{W/Hz}$ and $4.9\times10^{12}\,\text{W/Hz}$ for a dispersed burst and a narrowband signal at an SNR of 10.
Here the maximum effective area of GReX is used along with a $T_{sys}$ of $50\,\text{K}$.
The wide FoV of GReX places it as a useful tool in characterization of potential technosignatures enabling meaningful constraints in the observing band.  

\section{Deployment} \label{sec:deploy}

\begin{figure}[h!]
    \centering
    \includegraphics[width=\linewidth]{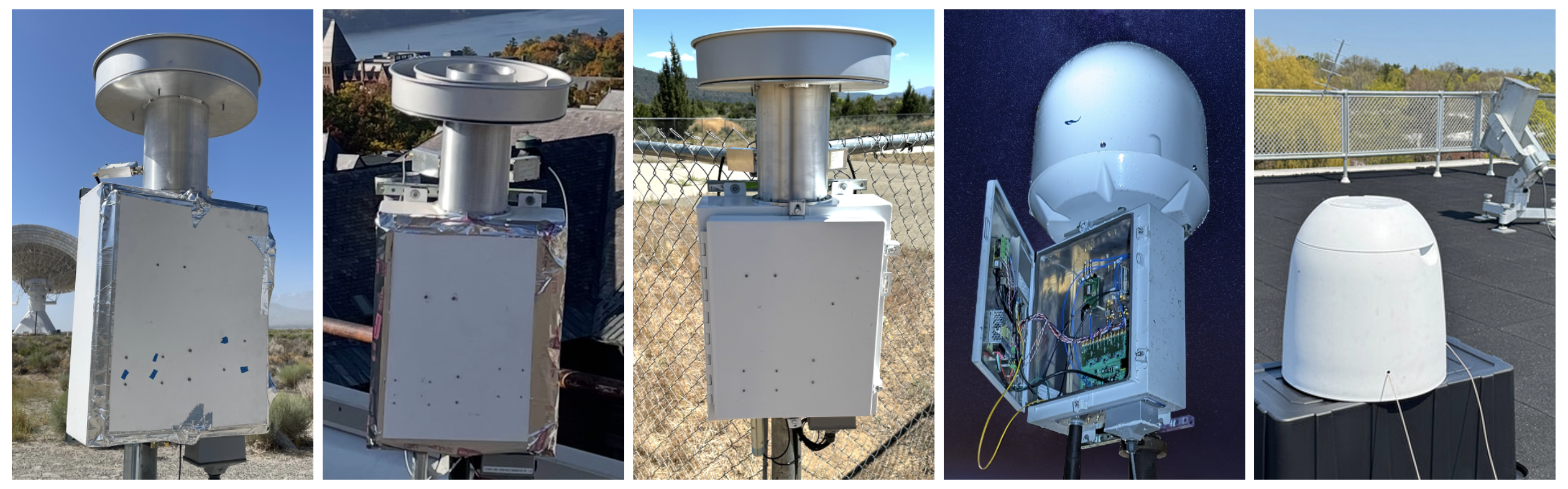}
    \caption{Deployed GReX terminals from left to right: Hat Creek Radio Observatory (USA), Cornell University Space Sciences Building (USA), Owens Valley Radio Observatory (USA), and Rosse Observatory (Ireland), and the Smithsonian Astrophysical Observatory at Harvard University (USA).}
    \label{fig:deployments}
\end{figure}

\begin{figure}[!h]
    \centering
    \includegraphics[width=\linewidth]{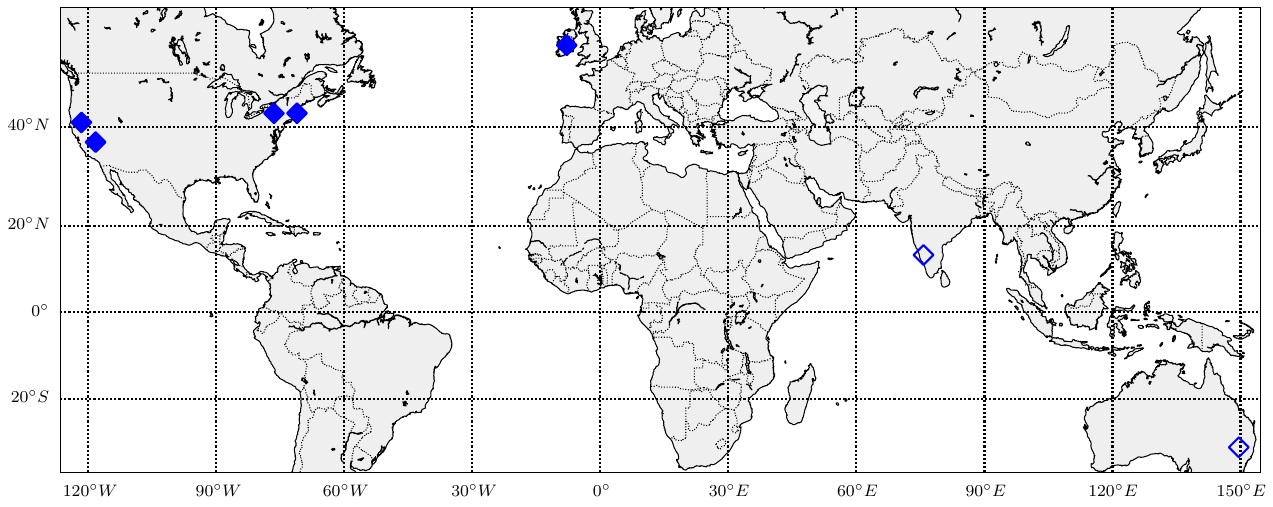}
    \caption{Mercator map with deployed and on-sky GReX stations ($\blacklozenge$) along with stations at various stages of deployment ($\lozenge$).}
    \label{fig:ground-deployments}
\end{figure}

\begin{figure}[!h]
    \centering
    \includegraphics[width=0.93\linewidth]{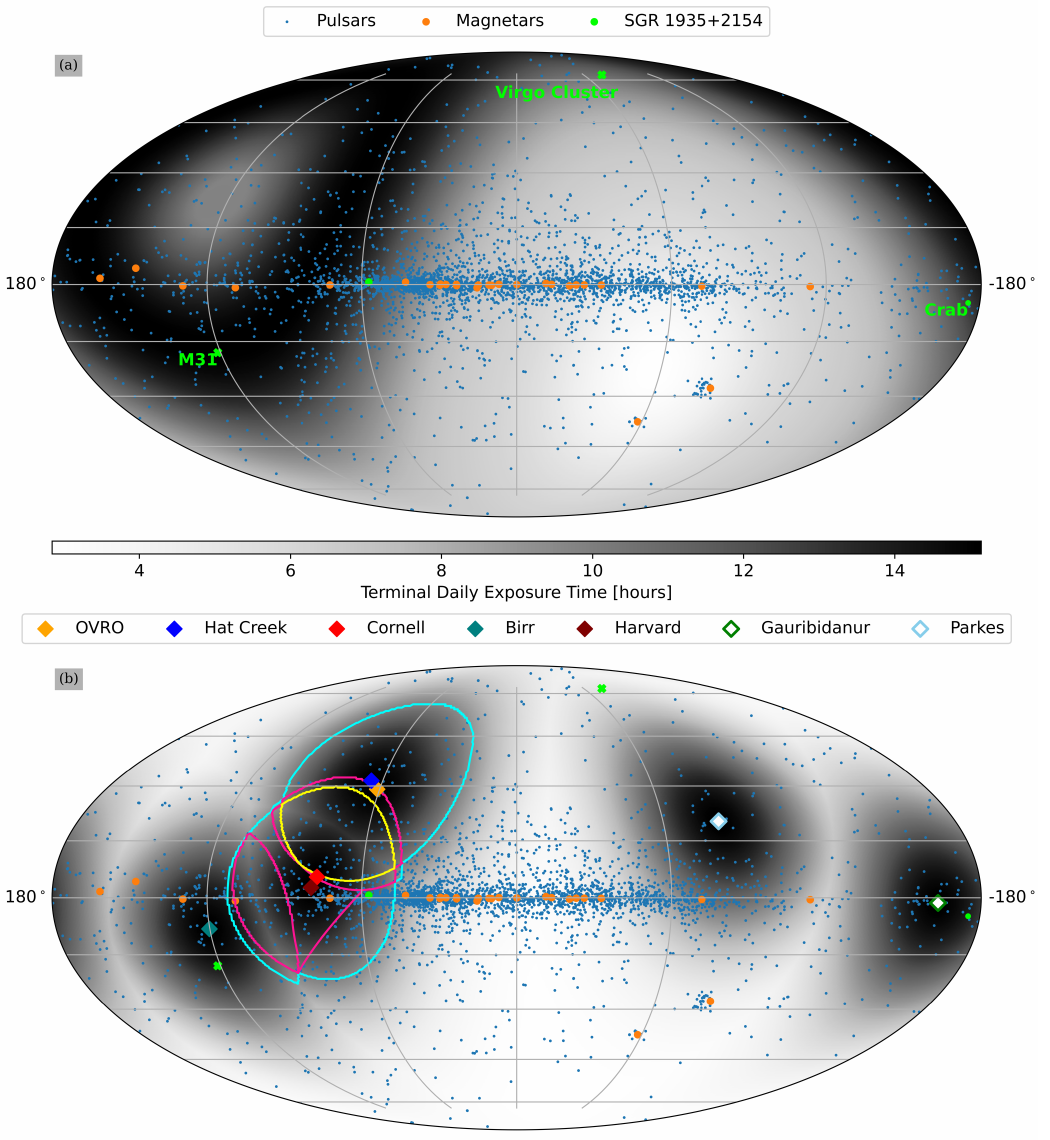}
    \caption{Figures showing current ($\blacklozenge$) and upcoming ($\lozenge$) GReX terminals (\autoref{fig:ground-deployments}), their coverage of the Galactic plane, and sources of interest. In \textbf{(a)} we show the total exposure time of the GReX system by averaging the simulated 2D Gaussian beam responses of each terminal over a single day. We include markers for the magnetar population taken from the McGill magnetar catelog \cite{McGillMagnetars} as that population produced the only detected Galactic FRB. We also include pulsars from the ATNF pulsar catalog \cite{PsrCat} as a proxy for the extent of the Galaxy. The Crab is singled out as giant pulses are potentially detectable with GReX. We also include M31 and the Virgo Cluster as locals from which an extraordinarily bright FRB might be detectable by GReX. In \textbf{(b)}, the grayscale heatmap represents the non-overlapping (joint) beam response for all terminals at a single instance in time, rather than the total exposure time over one day. We use this joint beam response to avoid overestimation from overlapping beams when calculating the exposure time in (a). The contours enclose regions of sky where the beams of multiple stations overlap, which we consider occurring when those beams are at or exceed their half-maximum gain. The cyan, pink, and yellow contours enclose regions with two, three, and four terminals satisfying this condition. Beams are simulated as 2D Gaussians with $\theta_\text{FWHM}=65.18^\circ$, in accordance with our determination of the beam response at $1420.4\,\text{MHz}$ in \autoref{sec:beam_model}.}
    \label{fig:sky-map}
\end{figure}

At the time of writing, five GReX units are currently operating on-sky (\autoref{fig:deployments}). 
Their locations and beam responses on-sky are shown in \autoref{fig:ground-deployments} and \autoref{fig:sky-map}, with two additional units planned for deployment at the Parkes Observatory in New South Wales, Australia and Gauribidanur, India.
The first operational unit was deployed at Owens Valley, followed by Cornell, Hat Creek, Birr, and Harvard.
Details of each deployment, including installation processes and site-specific considerations, are outlined below.

\subsection{Owens Valley Radio Observatory}
The deployment at the Owen's Valley Radio Observatory took the place of the previous STARE2 system in May 2024.
We made use of the same location and fiber optic connections.
Installing at the site involved the final assembly of the box, termination of fibers for the Ethernet connection, and testing.
Once the box was powered, we performed Y factor measurements to ensure we achieved our desired sensitivity.

\subsection{Cornell}
The installment of the GReX terminal at Cornell University followed a three-stage plan: site testing and preparation, debugging GReX hardware and software, and site installation with first-light measurements. Full deployment of the system was complete with the start of burst injections in early July 2024. 
The terminal has remained on-sky at this site since this initial deployment. 

Prior to receiving any hardware or software, we began a site-searching campaign to identify an optimal site to host the GReX terminal.
This site needs to meet four primary requirements: (1) easy access to electricity and internet, (2) a weatherproof and secure location to host the GReX server, (3) an unimpeded view of the sky from the terminal, and (4) a minimal level of radio frequency interference (RFI) within the observing band.
To facilitate this search, we developed a portable device for recording the RFI environment of a potential site. 
This device includes (i) a chargeable battery-pack for power, (ii) a Siglent Spectrum Analyzer for signal analysis, (iii) and a signal path consisting of a bias-T with 12-V power supply and a 1320-1580 MHz band pass filter\footnote{Mini-Circuits ZX75BP-1450-S+} all of which is contained within a (iv) portable rack case.
For consistency, we used the LNAs and cake pan antenna from the GReX system, with the initial goal of identifying any extreme continuous RFI that would immediately disqualify a potential site.
We then ran follow-up measurements spanning at least one day of collecting time at sites with acceptable RFI levels to search for the presence of any intermittent RFI that was missed during the first survey.
Following this second batch of measurements, we computed Y-factors and equivalent system temperatures (see \autoref{sec:sensitivity} for details) for each of the remaining potential host sites.
All sites showed roughly equivalent baseline system temperatures across the observing band. 
The Space Sciences Building (SSB) roof on Cornell's campus is the lowest system temperature site of those surveyed that meets all four primary site requirements. 
This, combined with the ease of access from having the GReX terminal located in the middle of campus, has led to us selecting the roof of SSB as the permanent Cornell GReX terminal location.

The second stage of deployment is characterized by the arrival of hardware (the server and GReX terminal) on site and subsequent software setup and bug-fixes. 
We received the servers and terminal box at Cornell at the end of June 2023. 
As the second deployed GReX terminal, we spent a few months working closely with team-members at Caltech on hardware and software bug fixes. 
This effort helped develop the comprehensive GReX software guide\footnote{\url{https://grex-telescope.github.io}} for ``ground up'' installation, from connecting the server to the Raspberry Pi and SNAP board in the terminal box to running the full GReX real-time data capture pipeline. 

With this bug-fixing period complete, we moved to setting up the terminal box on our selected site \autoref{fig:deployments} (Center Left) and taking first-light data collection. 
This included an initial collection of ADC RMS values and observation of the spectra collected on site. 
To prevent overflowing or underflowing of the ADCs, we tuned the adjustable gain in the terminal FEM. 
Two easy astrophysical signals to detect on first light are the presence of the solar continuum (seen as an overall increase in the spectrum as the sun enters the beam) and the HI emission line from Galactic hydrogen centered near $1420.4\,\text{MHz}$. 
The absence of either of these signals indicates that the terminal is not properly observing the sky and requires further fine-tuning, usually either due to an inappropriate FEM gain or spectral leakage within the LNAs. 
Once these checks are passed, we move on to carry out tests of the terminal sensitivity (\autoref{sec:sensitivity}) which will give us estimates of the receiver gain, the system temperature, and the resulting system equivalent flux density (SEFD). 

\subsection{Hat Creek Radio Observatory}

The GReX deployment at HCRO \autoref{fig:deployments} (Left) was completed in June 2024. 
Before installation, the box was first tested in a screened room using a spectrum analyzer connected to an omnidirectional antenna and external amplifier.
Measurements were conducted with the GReX unit turned off for baseline measurements and then powered on with the box both open and closed with RF absorbers placed inside the enclosure.
Steel wool was added to the fiber cable opening to enhance shielding.
These measurements (discussed in \autoref{sec:self_rfi}) show that the self-generated RFI is stronger than expected, and necessitated installation at a more remote location on-site to avoid interference with other experiments.

\subsection{Rosse Observatory}
The deployment of the GReX unit in Ireland was completed in December 2024 at the radio observatory of Trinity College Dublin.
The Rosse Observatory is situated in Birr, in the relatively low population density Irish Midlands.
The site already hosts the Irish LOw Frequency Array (LOFAR) station~\citep{lofar,ilofar} and other smaller experiments.
As the GReX deployment was in close proximity to the LOFAR station, any unintended emission from the GReX unit in the LOFAR band was measured in the lab, prior to deployment, see \autoref{sec:self_rfi}.
Due to the harsher weather conditions in Ireland, the GReX unit underwent further waterproofing with extra sealant on the box where all components are housed (see \autoref{fig:box_inside}).
The most notable modification is the use of a radome housing, as it is radio transparent and often used on marine vessels to protect equipment from harsh outdoor elements \autoref{fig:deployments} (Right).

\subsection{Self-Generated RFI} \label{sec:self_rfi}

\begin{figure}[h]
    \centering
    \includegraphics[width=\linewidth]{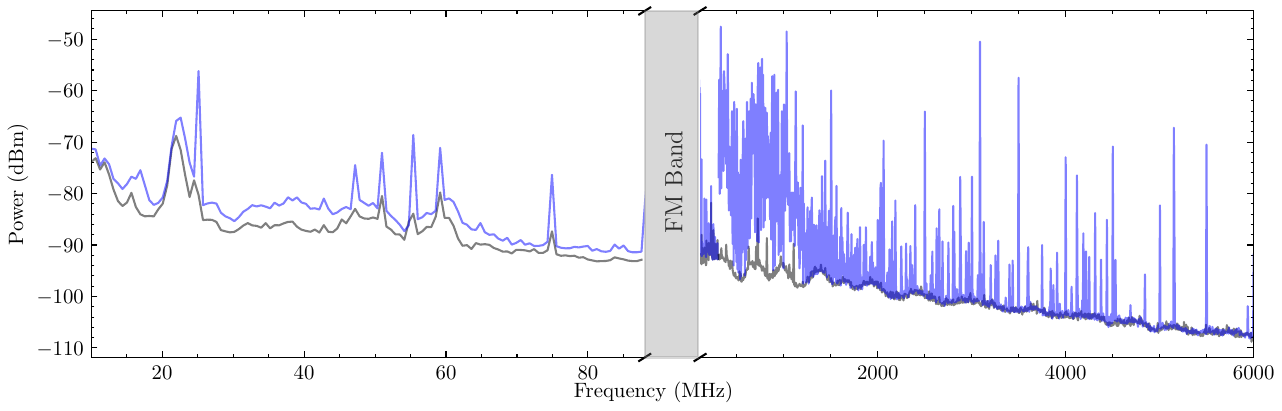}
    \caption{Measured power spectrum from $10-6000\,\text{MHz}$ of a closed GReX box.
    The blue trace shows the spectrum when the unit is powered on, while the gray trace represents the baseline measurement.
    The frequency range of $10-300\,\text{MHz}$ was measured at Birr in a lab and $300-6000\,\text{MHz}$ was measured at HCRO in a screened room.}
    \label{fig:RFISpectrum}
\end{figure}

\begin{figure}[h]
    \centering
    \includegraphics[width=\linewidth]{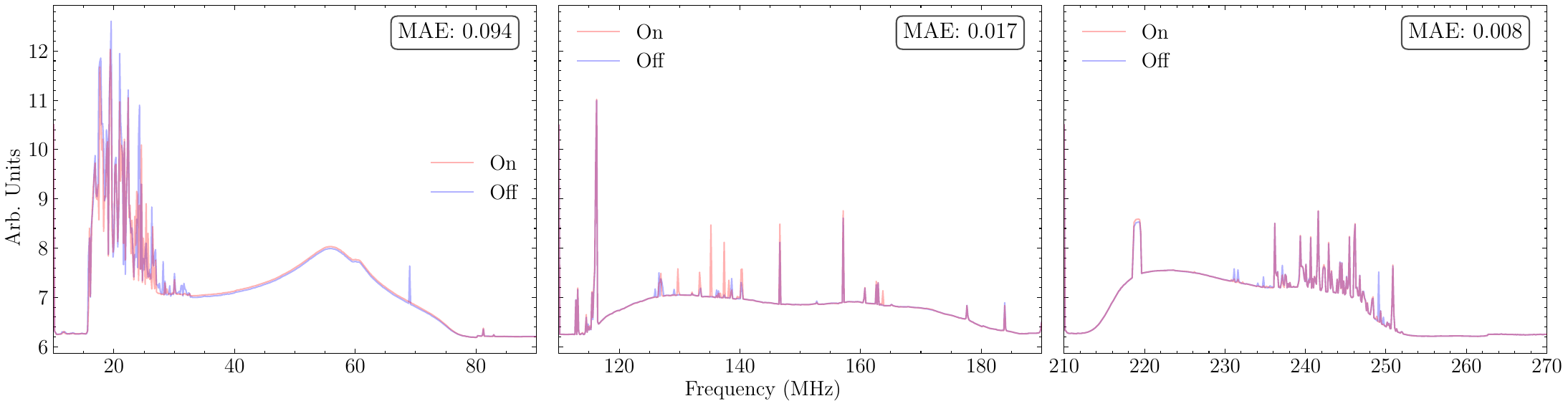}
    \caption{15-minute Stokes I spectra measured using I-LOFAR with the GReX unit powered on and off, covering all observing modes from $10-270\,\text{MHz}$. The mean absolute error (MAE) was calculated for each measurement, comparing the powered-on spectrum to the baseline (powered-off) measurement.}
    \label{fig:IE613-sst-test}
\end{figure}

As GReX is intended to be hosted at radio observatories as well as universities, tests of the self-generated RFI are critical to maintain spectrum purity at sensitive sites.
Lab tests of the box's emission were performed at HCRO and Birr, but not at OVRO due to lack of resources.
Additionally, Stokes I data were taken using I-LOFAR at Birr while GReX was operational.
The lab tests of the closed box's emission are shown in \autoref{fig:RFISpectrum}.
The results from the tests at HCRO in the $300-6000\,\text{MHz}$ band indicate that GReX in its current form is not yet suitable for installation close to sensitive telescopes observing in this band.
We believe the RFI retrofitting of the commercial enclosure used to house the electronics is primarily at fault.
Further experiments with other RFI gaskets and conductive tape show a marked improvement in radiated power, but not totally acceptable for sensitive sites, especially at L-band.
However, Stokes I data from I-LOFAR (shown in \autoref{fig:IE613-sst-test}) show no appreciable contamination.
Additionally, the various ongoing experiments at OVRO such as the DSA-110 and LWA have also not noticed any increase in local RFI.

While some mitigations have substantially improved the unit's radiated power, the box remains a moderate RFI source.
The self-generated RFI has not proven to be an issue for the FRB detection pipeline, nor has it interfered with experiments at operational radio observatories.
However, the emission remains an open issue and prevents installation at some sites.
Further work is needed to identify problematic components and improve shielding.

\section{Beam Modelling with Measurements of Galactic HI} \label{sec:beam_model}
The Hydrogen I (HI) emission line is an omnipresent and easily identifiable time-varying signal within the GReX observing band, making it an ideal metric for ensuring that a GReX terminal is observing the sky properly. 
Captured spectra in the \textit{Prometheus} database are accessed by Grafana to display estimates of the HI SNR in real time at the top of all Grafana dashboard monitors. 
While the HI-line intensity provides a useful sanity check that a terminal is properly observing the sky, its usefulness does not end there.
In this section, we present the results of our method for using the measured HI-line intensity at the OVRO and Cornell terminals to estimate the FWHM (and corresponding solid angle) of the GReX antenna beam-response.

\begin{figure}[!h]
    \centering    \includegraphics[width=1.0\textwidth]{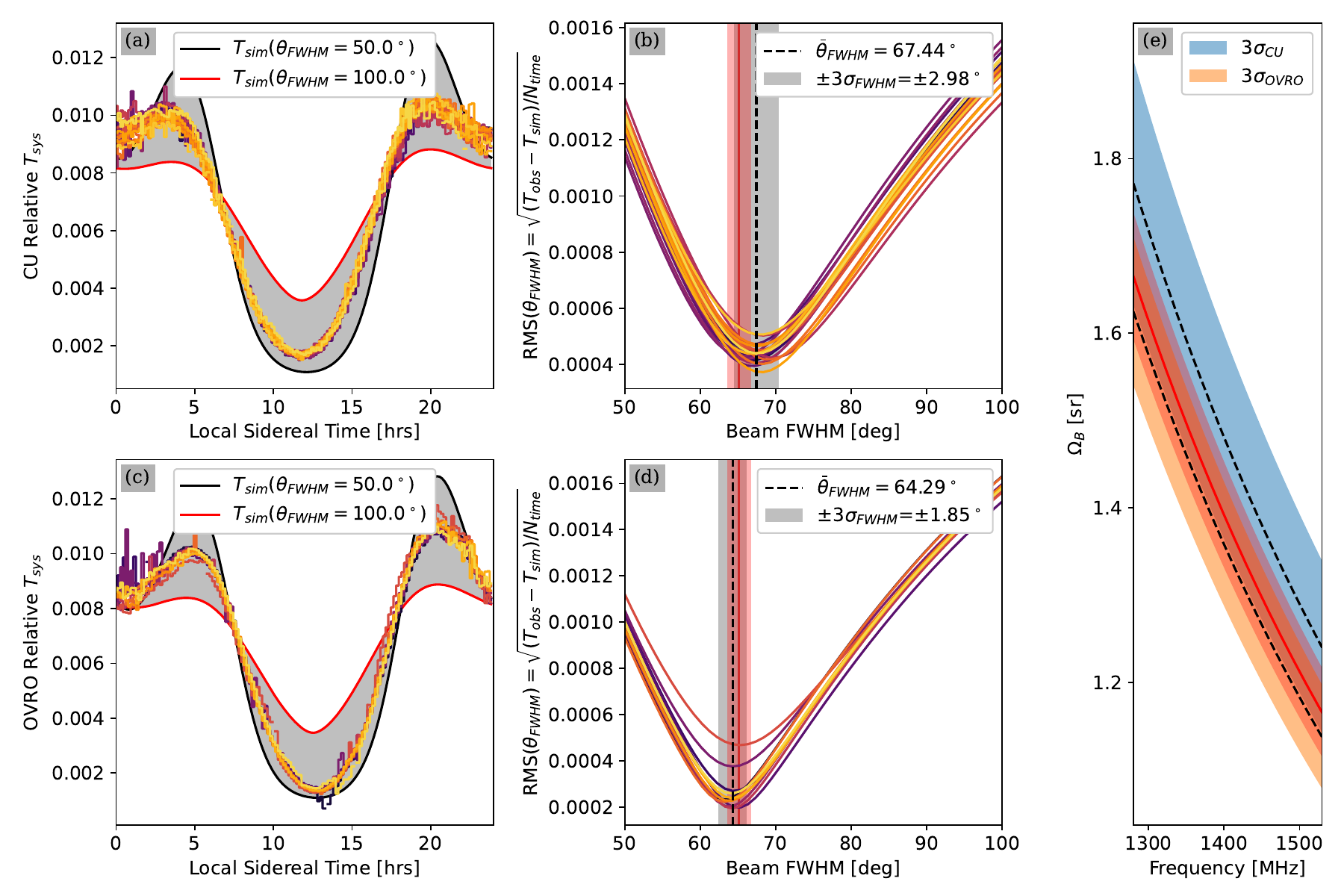}
    \caption{Observed data and corresponding simulations of the HI-line by the Cornell University (\textbf{(a)} and \textbf{(b)}) and OVRO (\textbf{(c)} and \textbf{(d)}) GReX terminals are used to fit the beam FWHM at 1420.4MHz. Simulations are rendered following the prescription in \autoref{sec:Asimulation}. The observed (colored step-function lines) and simulated data (gray shaded region bounded by red and black curves) are aligned by local sidereal time and samples overly affected by RFI are excluded when calculating the RMS error. The population of FWHM values that minimize the RMS errors for each terminal and each day are then used to determine the mean (black dashed line, \textbf{(b)} and \textbf{(d)}) and standard deviation of each terminal FWHM. The characteristic FWHM of a GReX terminal (\autoref{eq:popFWHM}) is presented in \textbf{(b)} and \textbf{(d)} as a solid red line with $3\sigma$ shading. We then scale the solid angle across the observing band according to the relationship $\Omega_B\propto f^{-2}$ in \textbf{(e)}.}\label{fig:HIsimFit}
\end{figure}

Following the prescription detailed in \autoref{sec:Asimulation}, we use the Leiden/Argentine/Bohn (LAB) all-sky survey of galactic neutral hydrogen \citep{LAB_survey_data} to simulate the expected antenna temperature as seen by the terminal over one sidereal day (sampled every 10 minutes) for fifty different beams with $\theta_\text{FWHM}$s ranging from $50^\circ$ to $100^\circ$. 
We query HI-line observations from the Cornell and OVRO \textit{Prometheus} databases with time steps of ten minutes over a month. 
We only keep days where the data is continuous across the day to consistently align the data and simulations according to the local sidereal time (LST) before normalizing both such that the average temperature over one day is a constant between all simulations and observations. 
For each time sample across the sidereal day, we calculate the observed HI-line mean intensity and standard deviation. 
Any samples that exceed a $3\sigma$ deviation from the mean are considered to be impacted by RFI and are ignored when calculating the root-mean squared (RMS) residuals between observations and models. 
Any days with more than 5\% of samples above this threshold are fully ignored. 
After RFI mitigation, there are 23 and 11 days remaining from the Cornell and OVRO measurements, respectively. 
With the data aligned, normalized, and cleaned of RFI, we calculate the RMS error between each day and each model. 
\autoref{fig:HIsimFit} presents the aligned and normalized observed and simulated HI-line (left) alongside the RMS errors of each day against all modeled beam-shapes (right).
While all RMS curves display consistent profiles which have minima that are clustered around the average FWHM, there are some curves in \autoref{fig:HIsimFit} \textbf{(d)} that minimize at a higher RMS than is typical. 
We attribute this to the presence of RFI-affected samples that did not get removed by our thresholding process.
The modeled $\theta_\text{FWHM}$s corresponding to the RMS minimum for each day are taken as the population of best-fit parameters. 
We determine the general best fit model as the mean of this population, with a $1\sigma$ uncertainty in the FWHM given by the population standard deviation. 
Using this method, we determine that the Cornell terminal has a beam response at $1420.4\,\text{MHz}$ that is well-characterized as a 2D Gaussian with $\theta_\text{FWHM}^\text{CU}(1420.4\,\text{MHz})=67.44^\circ\pm0.99^\circ$. 
The OVRO terminal is similarly described by a $\theta_\text{FWHM}^\text{OVRO}(1420.4\,\text{MHz})=64.29^\circ\pm0.62^\circ$. 

Since all GReX terminals are outfitted with identical feeds, we consider the beam response of a typical feed to be a good description for all feeds. 
This typical beam response at $1420.4\,\text{MHz}$ is then characterized as
\begin{equation}
    \theta_\text{FWHM} = \frac{\frac{\theta_\text{FWHM}^\text{OVRO}}{\sigma_\text{OVRO}^2}+\frac{\theta_\text{FWHM}^\text{CU}}{\sigma_\text{CU}^2}}{\frac{1}{\sigma_\text{OVRO}^2}+\frac{1}{\sigma_\text{CU}^2}}\pm\sqrt{\frac{1}{\frac{1}{\sigma_\text{OVRO}^2}+\frac{1}{\sigma_\text{CU}^2}}}=65.18^\circ\pm0.53\label{eq:popFWHM}
\end{equation}
with corresponding solid angle at that frequency of \begin{equation}
    \Omega_B=\int_\Omega d\Omega \exp{\left[-4\ln2\left(\frac{\theta}{\theta_\text{FWHM}}\right)^2\right]}= 1.36\,\pm0.02\,\text{sr}.
    \label{eq:beam_solid_angle}
\end{equation}
We use $\Omega_B\propto f^{-2}$ to determine the solid angle as a function of frequency,
\begin{equation}
\Omega_B(f)=\frac{1931.7\pm28.4}{f^2} \,\text{sr}\,\text{MHz}^2.
\end{equation}
This functional relationship (and equivalent comparisons for the individual terminal fits) is presented in \autoref{fig:HIsimFit} (\textbf{e}). 
We determine from this relation that the beam solid angle near our central frequency is $\Omega_B(1400\,\text{MHz})=1.40\,\pm0.02\,\text{sr}$. 
We also report the maximum effective area $A_0$ of the receiver across the band as $A_0=328\,\pm5\,\text{cm}^2$ according to \autoref{eq:Ae}. 
\section{Sensitivity} \label{sec:sensitivity}
Understanding the sensitivity of the GReX instrument to incoming signals is necessary for defining the strength of those signals and being able to place them in an astrophysical context.

\begin{figure}[h!]
    \centering
    \includegraphics[width=1.0\textwidth]{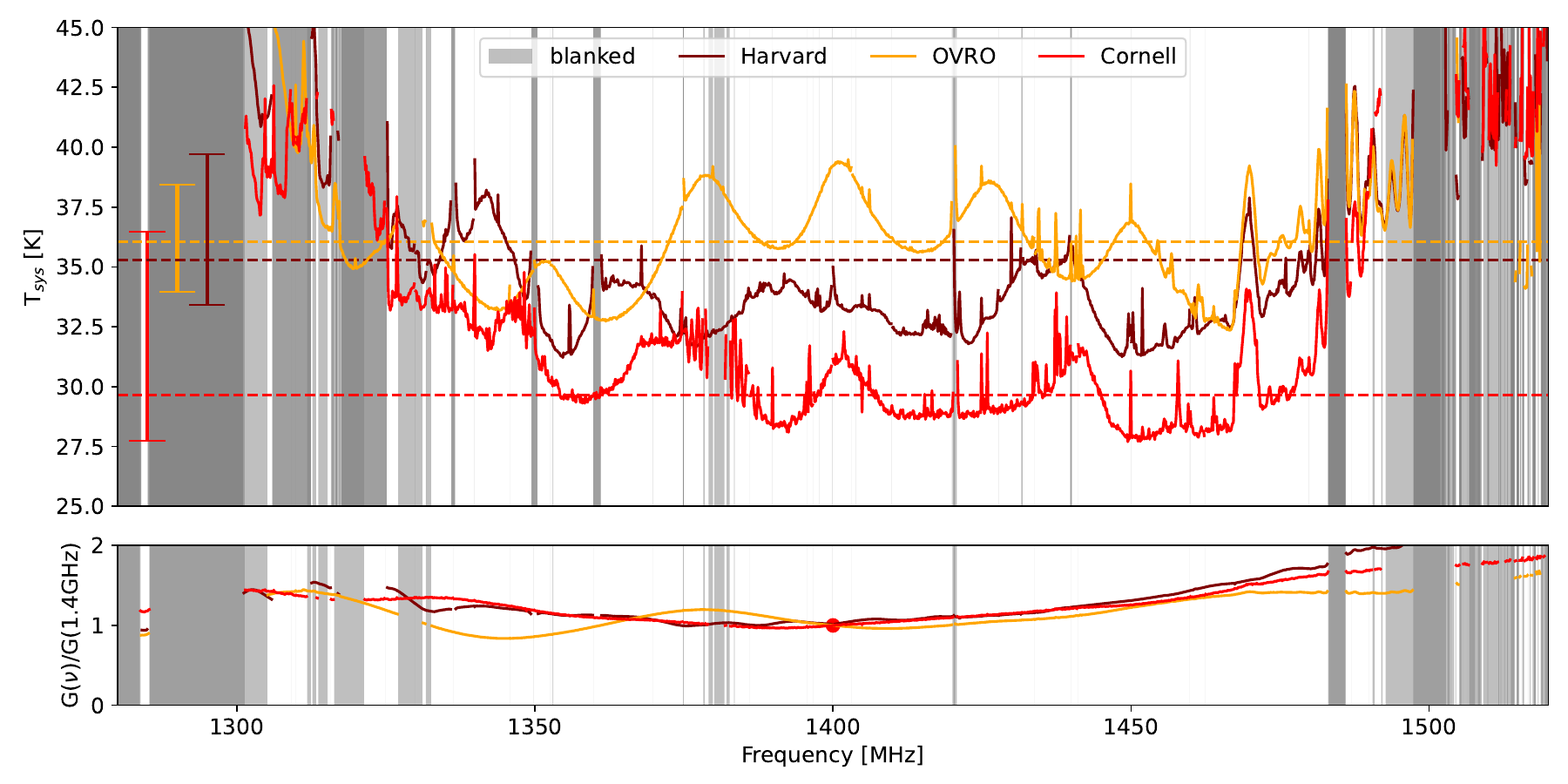}
    \caption{System temperature and relative receiver gain measured for the GReX terminals at Cornell, OVRO, and Harvard.
    Dashed lines show the median system temperature with associated $16^{th}$ and $84^{th}$ quantile error bars.
    }
    \label{fig:Tsys}
\end{figure}

The receiver gains and the system temperatures for the Cornell, OVRO, and Harvard stations are shown in \autoref{fig:Tsys}, computed using a Y-factor test (\autoref{eq:Trec}).
Sky measurements are taken before daybreak or after sundown to ensure emission from the sun does not enter the beam. 
This allows us to assume a constant sky temperature of $5.5\,\text{K}$, with $2.7\,\text{K}$ from the CMB, $1.9\,\text{K}$ from atmospheric effects, and $0.9\,\text{K}$ from galactic effects \citep{SkyTemp-Vine}. 
We placed a slab of ambient-temperature (nominally $290\,\text{K}\pm5\,\text{K}$) radio-absorbent foam over the terminal antenna while taking hot data for the Y-factor test.
Data were saved in the Stokes I filterbank format for the hot foam and cold sky states.

The system noise temperature for a GReX terminal is approximately $35\,\text{K}$, averaged across the band.
Obstructions in the beam such as buildings and trees will worsen the noise, as well as the presence of strong RFI desensitizing the receiver.
As such, optimal placement and configuration of attenuation/gains is critical to maximizing performance.

The second essential measure of system sensitivity is the forward gain ($g_f$) of the receiver. 
This quantity depends primarily on the integrated beam response function across the sky (\autoref{eq:FG}, \autoref{eq:Ae}), since that is the value mediating between the true brightness temperature integrated across the telescope beam and the nominal effective temperature representing the observed temperature (\autoref{eq:Teff}, \autoref{eq:Bnorm}).
Applying the effective area of the terminal found in \autoref{sec:beam_model} to \autoref{eq:FG}, we compute the forward gain as $g_f=84.2\,\text{kJy}\,\text{K}^{-1}\pm1.3\,\text{kJy}\,\text{K}^{-1}$.
Multiplying this value by the system temperature gives an SEFD of approximately $3\,\text{MJy}$.
When determining our detection threshold, we consider a proto-typical burst that fills the band and lasts $1\,\text{ms}$.
Our detectors are sensitive to an effective bandwidth of $188\,\text{MHz}$ ($75\%$ of the total bandwidth) and the use of matched-filtering with a set of boxcars in the detection pipeline allows our integration time to match the $1\,\text{ms}$ burst width. 
By averaging over these samples in frequency, time, and both polarizations, we find that for an event to attain an SNR of 10 (chosen as a fiducial value with a reasonable empirically-determined false-positive rate), it must have a corresponding flux density of at least $\sim50\,\text{kJy}$. 

This is a minimum requirement, as the angular deviation of a source away from the center of the detector beam at the time of observation increases the required flux density necessary to detect the burst.

\section{Upper Limit of Bright FRBs} \label{sec:rate}

We constrain the upper limit on FRBs at least as bright as our detection threshold by estimating the number of detections over some time as a Poissonian process given the single detection by STARE2 and, at present, zero detections by GReX.
For this analysis, we will only consider the time on sky at the OVRO site.
The Poissonian process has a probability distribution of
\begin{equation}
    p(k|\lambda,\tau) = \frac{(\lambda\tau)^ke^{-\lambda\tau}}{k!}\,,
\end{equation}
for $k$ events given the event rate $\lambda$ and observation time $\tau$.
To estimate the event rate $\lambda$ with $k=1$ after time on sky of $\tau_1$, we solve for $p(\lambda|k=1,\tau_1)$ via Bayes' theorem assuming a flat prior on $\lambda$ where $p(\lambda)\propto 1$ via
\begin{equation}
    p(\lambda|k=1,\tau_1) = \frac{p(k=1|\lambda,\tau_1)p(\lambda)}{p(k=1)}=\frac{\lambda\tau_1e^{-\lambda\tau_1}}{\int_{0}^{\infty}d\lambda \lambda\tau_1e^{-\lambda\tau_1}} = \tau_1^2\lambda e^{-\lambda\tau_1}.
    \label{eq:k1_p}
\end{equation}
This result is a gamma distribution with a scale of 2 and a rate parameter of $\tau_1$.

Given our center-frequency beam solid angle of $\sim1.5\,\text{sr}$ from \autoref{eq:beam_solid_angle}, at any given time the station is sensitive to only 12\% of the sky.
At the time of the STARE2 detection of FRB 200428, it had been observing for 448 days.
Assuming a 75\% observational duty-cycle, we compute $\tau_1 = 1.23\,\text{yr}\,\times\,0.12\,\text{sky}\,\times\,0.75=0.11\,\text{sky yr}$.
STARE2 continued to observe after FRB 200428 until it was superseded by GReX in May 2024.
In total, a station at OVRO has now been observing for 6.3 years.
We then compute a new rate parameter of $\tau_1 = 6.3\,\text{yr}\,\times\,0.12\,\text{sky}\,\times\,0.75=0.567\,\text{sky yr}$.

\begin{figure}[ht!]
    \centering
    \includegraphics[width=0.8\linewidth]{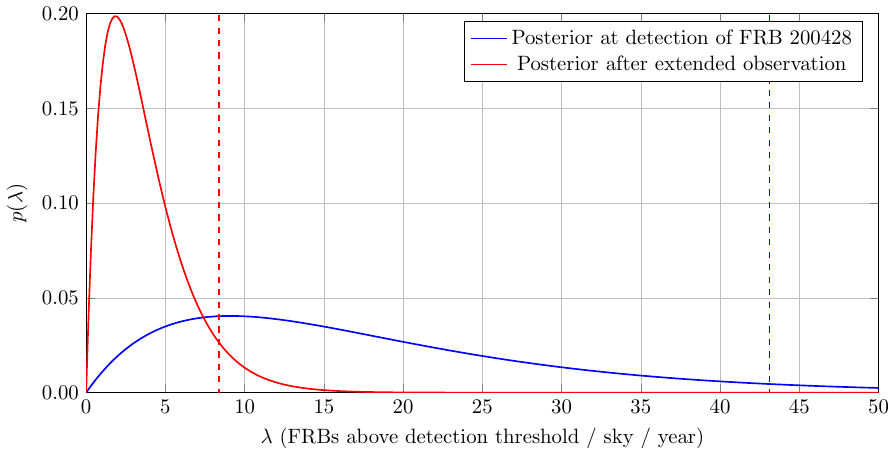}
    \caption{Posterior distribution of FRB event rate $\lambda$ with associated dashed 95\% confidence upper limits}
    \label{fig:rate_pdfs}
\end{figure}

Shown in \autoref{fig:rate_pdfs} are the probability distribution functions for the event rate $\lambda$, implied by the observation time at the detection of FRB 200428 and the current total observation time.
The STARE2 detection placed an upper limit on the rate parameter of $43.13\,\text{sky}^{-1}\,\text{yr}^{-1}$, with wide uncertainty due to the limited time on sky.
Adding the subsequent non-detection time significantly reduces this estimate to $8.37\,\text{sky}^{-1}\,\text{yr}^{-1}$.
A more thorough analysis of the rate including overlapping sky and time, luminosity, etc. will follow in a subsequent paper.

\section{Conclusions and Outlook} \label{sec:conclusions}
We have described the Galactic Radio Explorer (GReX), a network of low-cost $1.4\,\text{GHz}$ radio antennas searching for bright transients between $32.768\,\mu\text{s}$ and $1.024\,\text{ms}$.
Terminals have been deployed at multiple sites around the world, though most are currently in the US.
Since the discovery of FRB\,200428, GReX and an upgraded STARE2 have more than quadrupled the total square degree hours devoted to ultra-bright FRB searching.
The improved system temperature of GReX LNAs over the original STARE2 hardware results in $\sim$\,twice the sensitivity, resulting in a deeper search.
We also search for extremely narrow bursts $\ll1\,\text{ms}$, which is not possible at lower frequencies or in systems with a large number of beams.
With no detections, we update the all-sky rate of FRBs above $\sim50\,\text{kJy}$ to be no larger than 8.37\,sky$^{-1}$\,yr$^{-1}$ at 95$\%$ confidence.
This indicates that FRB-like emission from Galactic sources is rare, and FRB\,200428 was even more unusual than initially thought. 

The GReX boxes have proven to be replicable and stable.
For example, the OVRO GReX terminal did not require on-site human intervention for more than a year of observation. Five sites are on-sky and continuously searching for bright FRBs.
These locations are  OVRO in California, the Cornell GReX in Ithaca, New York, Harvard GReX in Cambridge Massachusetts, Hat Creek GReX in California, and an Ireland station in Birr.
A box has been shipped and tested in New South Wales, though not yet deployed.
In the future, we hope to expand to multiple sites in the Southern Hemisphere to improve exposure to the Galactic plane, where most magnetars reside. 

\begin{acknowledgments}
The authors would like to thank Courtney Keeler and David Hodge for all the assistance constructing, testing, and shipping the boxes.
They would also like to acknowledge the staff at OVRO, HCRO, Rosse Observatory, and CSIRO for their assistance testing and working on deploying these instruments on site.
The Rosse Observatory, in Birr, is operated by Trinity College Dublin.
EFK acknowledges that the GReX unit in Birr was made possible thanks to a grant from the Trinity College Dublin Association and Trust.
OAJ acknowledges the support of Breakthrough Listen, which is managed by the Breakthrough Prize Foundation. 
Support for the construction of GReX was provided by Mt. Cuba Astronomical Foundation and a private donation from Terry Cisco.
\end{acknowledgments}

\vspace{5mm}

\software{GNU Parallel~\citep{tange_2025_14911163}, Astropy~\citep{astropy:2013, astropy:2018, astropy:2022}, HEALPix~\citep{HEALPix}, PSRDADA~\citep{vanstratenPSRDADADistributedAcquisition2021}, HEIMDALL~\citep{barsdellAcceleratingIncoherentDedispersion2012}}

\bibliography{main}{}

\begin{thebibliography}{}
\expandafter\ifx\csname natexlab\endcsname\relax\def\natexlab#1{#1}\fi
\providecommand{\url}[1]{\href{#1}{#1}}
\providecommand{\dodoi}[1]{doi:~\href{http://doi.org/#1}{\nolinkurl{#1}}}
\providecommand{\doeprint}[1]{\href{http://ascl.net/#1}{\nolinkurl{http://ascl.net/#1}}}
\providecommand{\doarXiv}[1]{\href{https://arxiv.org/abs/#1}{\nolinkurl{https://arxiv.org/abs/#1}}}

\bibitem[{{Astropy Collaboration} {et~al.}(2013){Astropy Collaboration},
  {Robitaille}, {Tollerud}, {Greenfield}, {Droettboom}, {Bray}, {Aldcroft},
  {Davis}, {Ginsburg}, {Price-Whelan}, {Kerzendorf}, {Conley}, {Crighton},
  {Barbary}, {Muna}, {Ferguson}, {Grollier}, {Parikh}, {Nair}, {Unther},
  {Deil}, {Woillez}, {Conseil}, {Kramer}, {Turner}, {Singer}, {Fox}, {Weaver},
  {Zabalza}, {Edwards}, {Azalee Bostroem}, {Burke}, {Casey}, {Crawford},
  {Dencheva}, {Ely}, {Jenness}, {Labrie}, {Lim}, {Pierfederici}, {Pontzen},
  {Ptak}, {Refsdal}, {Servillat}, \& {Streicher}}]{astropy:2013}
{Astropy Collaboration}, {Robitaille}, T.~P., {Tollerud}, E.~J., {et~al.} 2013,
  \aap, 558, A33, \dodoi{10.1051/0004-6361/201322068}

\bibitem[{{Astropy Collaboration} {et~al.}(2018){Astropy Collaboration},
  {Price-Whelan}, {Sip{\H{o}}cz}, {G{\"u}nther}, {Lim}, {Crawford}, {Conseil},
  {Shupe}, {Craig}, {Dencheva}, {Ginsburg}, {Vand erPlas}, {Bradley},
  {P{\'e}rez-Su{\'a}rez}, {de Val-Borro}, {Aldcroft}, {Cruz}, {Robitaille},
  {Tollerud}, {Ardelean}, {Babej}, {Bach}, {Bachetti}, {Bakanov}, {Bamford},
  {Barentsen}, {Barmby}, {Baumbach}, {Berry}, {Biscani}, {Boquien}, {Bostroem},
  {Bouma}, {Brammer}, {Bray}, {Breytenbach}, {Buddelmeijer}, {Burke},
  {Calderone}, {Cano Rodr{\'\i}guez}, {Cara}, {Cardoso}, {Cheedella}, {Copin},
  {Corrales}, {Crichton}, {D'Avella}, {Deil}, {Depagne}, {Dietrich}, {Donath},
  {Droettboom}, {Earl}, {Erben}, {Fabbro}, {Ferreira}, {Finethy}, {Fox},
  {Garrison}, {Gibbons}, {Goldstein}, {Gommers}, {Greco}, {Greenfield},
  {Groener}, {Grollier}, {Hagen}, {Hirst}, {Homeier}, {Horton}, {Hosseinzadeh},
  {Hu}, {Hunkeler}, {Ivezi{\'c}}, {Jain}, {Jenness}, {Kanarek}, {Kendrew},
  {Kern}, {Kerzendorf}, {Khvalko}, {King}, {Kirkby}, {Kulkarni}, {Kumar},
  {Lee}, {Lenz}, {Littlefair}, {Ma}, {Macleod}, {Mastropietro}, {McCully},
  {Montagnac}, {Morris}, {Mueller}, {Mumford}, {Muna}, {Murphy}, {Nelson},
  {Nguyen}, {Ninan}, {N{\"o}the}, {Ogaz}, {Oh}, {Parejko}, {Parley}, {Pascual},
  {Patil}, {Patil}, {Plunkett}, {Prochaska}, {Rastogi}, {Reddy Janga},
  {Sabater}, {Sakurikar}, {Seifert}, {Sherbert}, {Sherwood-Taylor}, {Shih},
  {Sick}, {Silbiger}, {Singanamalla}, {Singer}, {Sladen}, {Sooley},
  {Sornarajah}, {Streicher}, {Teuben}, {Thomas}, {Tremblay}, {Turner},
  {Terr{\'o}n}, {van Kerkwijk}, {de la Vega}, {Watkins}, {Weaver}, {Whitmore},
  {Woillez}, {Zabalza}, \& {Astropy Contributors}}]{astropy:2018}
{Astropy Collaboration}, {Price-Whelan}, A.~M., {Sip{\H{o}}cz}, B.~M., {et~al.}
  2018, \aj, 156, 123, \dodoi{10.3847/1538-3881/aabc4f}

\bibitem[{{Astropy Collaboration} {et~al.}(2022){Astropy Collaboration},
  {Price-Whelan}, {Lim}, {Earl}, {Starkman}, {Bradley}, {Shupe}, {Patil},
  {Corrales}, {Brasseur}, {N{"o}the}, {Donath}, {Tollerud}, {Morris},
  {Ginsburg}, {Vaher}, {Weaver}, {Tocknell}, {Jamieson}, {van Kerkwijk},
  {Robitaille}, {Merry}, {Bachetti}, {G{"u}nther}, {Aldcroft},
  {Alvarado-Montes}, {Archibald}, {B{'o}di}, {Bapat}, {Barentsen}, {Baz{'a}n},
  {Biswas}, {Boquien}, {Burke}, {Cara}, {Cara}, {Conroy}, {Conseil}, {Craig},
  {Cross}, {Cruz}, {D'Eugenio}, {Dencheva}, {Devillepoix}, {Dietrich},
  {Eigenbrot}, {Erben}, {Ferreira}, {Foreman-Mackey}, {Fox}, {Freij}, {Garg},
  {Geda}, {Glattly}, {Gondhalekar}, {Gordon}, {Grant}, {Greenfield}, {Groener},
  {Guest}, {Gurovich}, {Handberg}, {Hart}, {Hatfield-Dodds}, {Homeier},
  {Hosseinzadeh}, {Jenness}, {Jones}, {Joseph}, {Kalmbach}, {Karamehmetoglu},
  {Ka{l}uszy{'n}ski}, {Kelley}, {Kern}, {Kerzendorf}, {Koch}, {Kulumani},
  {Lee}, {Ly}, {Ma}, {MacBride}, {Maljaars}, {Muna}, {Murphy}, {Norman},
  {O'Steen}, {Oman}, {Pacifici}, {Pascual}, {Pascual-Granado}, {Patil},
  {Perren}, {Pickering}, {Rastogi}, {Roulston}, {Ryan}, {Rykoff}, {Sabater},
  {Sakurikar}, {Salgado}, {Sanghi}, {Saunders}, {Savchenko}, {Schwardt},
  {Seifert-Eckert}, {Shih}, {Jain}, {Shukla}, {Sick}, {Simpson},
  {Singanamalla}, {Singer}, {Singhal}, {Sinha}, {Sip{H{o}}cz}, {Spitler},
  {Stansby}, {Streicher}, {{{S}}umak}, {Swinbank}, {Taranu}, {Tewary},
  {Tremblay}, {Val-Borro}, {Van Kooten}, {Vasovi{'c}}, {Verma}, {de Miranda
  Cardoso}, {Williams}, {Wilson}, {Winkel}, {Wood-Vasey}, {Xue}, {Yoachim},
  {Zhang}, {Zonca}, \& {Astropy Project Contributors}}]{astropy:2022}
{Astropy Collaboration}, {Price-Whelan}, A.~M., {Lim}, P.~L., {et~al.} 2022,
  \apj, 935, 167, \dodoi{10.3847/1538-4357/ac7c74}

\bibitem[{Barsdell {et~al.}(2012)Barsdell, Bailes, Barnes, \&
  Fluke}]{barsdellAcceleratingIncoherentDedispersion2012}
Barsdell, B.~R., Bailes, M., Barnes, D.~G., \& Fluke, C.~J. 2012, Monthly
  Notices of the Royal Astronomical Society, 422, 379,
  \dodoi{10.1111/j.1365-2966.2012.20622.x}

\bibitem[{Bochenek {et~al.}(2020)Bochenek, McKenna, Belov, Kocz, Kulkarni,
  Lamb, Ravi, \& Woody}]{bochenekSTARE2DetectingFast2020}
Bochenek, C.~D., McKenna, D.~L., Belov, K.~V., {et~al.} 2020, Publications of
  the Astronomical Society of the Pacific, 132, 034202,
  \dodoi{10.1088/1538-3873/ab63b3}

\bibitem[{{Caleb} {et~al.}(2024){Caleb}, {Lenc}, {Kaplan}, {Murphy}, {Men},
  {Shannon}, {Ferrario}, {Rajwade}, {Clarke}, {Giacintucci}, {Hurley-Walker},
  {Hyman}, {Lower}, {McSweeney}, {Ravi}, {Barr}, {Buchner}, {Flynn}, {Hessels},
  {Kramer}, {Pritchard}, \& {Stappers}}]{Caleb}
{Caleb}, M., {Lenc}, E., {Kaplan}, D.~L., {et~al.} 2024, Nature Astronomy, 8,
  1159, \dodoi{10.1038/s41550-024-02277-w}

\bibitem[{CHIME/FRB~Collaboration {et~al.}(2020)CHIME/FRB~Collaboration,
  Bandura, Bhardwaj, Bij, Boyce, Boyle, Brar, Cassanelli, Chawla, Chen, Cliche,
  Cook, Cubranic, Curtin, Denman, Dobbs, Dong, Fandino, Fonseca, Gaensler,
  Giri, Good, Halpern, Hill, Hinshaw, Höfer, Josephy, Kania, Kaspi, Landecker,
  Leung, Li, Lin, Masui, Mckinven, Mena-Parra, Merryfield, Meyers, Michilli,
  Milutinovic, Mirhosseini, Münchmeyer, Naidu, Newburgh, Ng, Patel, Pen,
  Pinsonneault-Marotte, Pleunis, Quine, Rafiei-Ravandi, Rahman, Ransom, Renard,
  Sanghavi, Scholz, Shaw, Shin, Siegel, Singh, Smegal, Smith, Stairs, Tan,
  Tendulkar, Tretyakov, Vanderlinde, Wang, Wulf, Zwaniga, \& {The CHIME/FRB
  Collaboration}}]{andersen_bright_2020}
CHIME/FRB~Collaboration, Andersen, B., Bandura, K., Bhardwaj, M., {et~al.}
  2020, Nature, 587, 54, \dodoi{10.1038/s41586-020-2863-y}

\bibitem[{Connor {et~al.}(2021)Connor, Shila, Kulkarni, Flygare, Hallinan, Li,
  Lu, Ravi, \& Weinreb}]{connorGalacticRadioExplorer2021a}
Connor, L., Shila, K.~A., Kulkarni, S.~R., {et~al.} 2021, Publications of the
  Astronomical Society of the Pacific, 133, 075001,
  \dodoi{10.1088/1538-3873/ac0bcc}

\bibitem[{{Cordes} \& {Chatterjee}(2019)}]{cordes-review}
{Cordes}, J.~M., \& {Chatterjee}, S. 2019, \araa, 57, 417,
  \dodoi{10.1146/annurev-astro-091918-104501}

\bibitem[{{Eftekhari} {et~al.}(2025){Eftekhari}, {Dong}, {Fong}, {Shah},
  {Simha}, {Andersen}, {Andrew}, {Bhardwaj}, {Cassanelli}, {Chatterjee},
  {Coulter}, {Fonseca}, {Gaensler}, {Gordon}, {Hessels}, {Ibik}, {Joseph},
  {Kahinga}, {Kaspi}, {Kharel}, {Kilpatrick}, {Lanman}, {Lazda}, {Leung},
  {Liu}, {Mas-Ribas}, {Masui}, {Mckinven}, {Mena-Parra}, {Miller}, {Nimmo},
  {Pandhi}, {Patil}, {Pearlman}, {Pleunis}, {Prochaska}, {Rafiei-Ravandi},
  {Sammons}, {Scholz}, {Shin}, {Smith}, \& {Stairs}}]{chime_elliptical_1}
{Eftekhari}, T., {Dong}, Y., {Fong}, W., {et~al.} 2025, \apjl, 979, L22,
  \dodoi{10.3847/2041-8213/ad9de2}

\bibitem[{Eriksson {et~al.}(2013)Eriksson, H{\"a}ggstr{\"o}m, Aittamaa,
  Kruglyak, \& Lindgren}]{erikssonRealtimeMassesStep2013}
Eriksson, J., H{\"a}ggstr{\"o}m, F., Aittamaa, S., Kruglyak, A., \& Lindgren,
  P. 2013, in 2013 8th {{IEEE International Symposium}} on {{Industrial
  Embedded Systems}} ({{SIES}}), 110--113

\bibitem[{Gajjar {et~al.}(2022)Gajjar, LeDuc, Chen, Siemion, Sheikh, Brzycki,
  Croft, Czech, DeBoer, DeMarines, Drew, Isaacson, Lacki, Lebofsky, MacMahon,
  Ng, de~Pater, Perez, Price, Suresh, Webb, \& Worden}]{gajjar_searching_2022}
Gajjar, V., LeDuc, D., Chen, J., {et~al.} 2022, The Astrophysical Journal, 932,
  81, \dodoi{10.3847/1538-4357/ac6dd5}

\bibitem[{{G{\'o}rski} {et~al.}(2005){G{\'o}rski}, {Hivon}, {Banday},
  {Wandelt}, {Hansen}, {Reinecke}, \& {Bartelmann}}]{HEALPix}
{G{\'o}rski}, K.~M., {Hivon}, E., {Banday}, A.~J., {et~al.} 2005, \apj, 622,
  759, \dodoi{10.1086/427976}

\bibitem[{Hickish {et~al.}(2016)Hickish, Abdurashidova, Ali, Buch, Chaudhari,
  Chen, Dexter, Domagalski, Ford, Foster, George, Greenberg, Greenhill,
  Isaacson, Jiang, Jones, Kapp, Kriel, Lacasse, Lutomirski, MacMahon, Manley,
  Martens, McCullough, Muley, New, Parsons, Price, Primiani, Ray, Siemion,
  Van~Tonder, Vertatschitsch, Wagner, Weintroub, \&
  Werthimer}]{hickishDecadeDevelopingRadioAstronomy2016a}
Hickish, J., Abdurashidova, Z., Ali, Z., {et~al.} 2016, Journal of Astronomical
  Instrumentation, 05, 1641001, \dodoi{10.1142/S2251171716410014}

\bibitem[{{Hurley-Walker} {et~al.}(2022){Hurley-Walker}, {Zhang}, {Bahramian},
  {McSweeney}, {O'Doherty}, {Hancock}, {Morgan}, {Anderson}, {Heald}, \&
  {Galvin}}]{HurleyWalker}
{Hurley-Walker}, N., {Zhang}, X., {Bahramian}, A., {et~al.} 2022, \nat, 601,
  526, \dodoi{10.1038/s41586-021-04272-x}

\bibitem[{Johnson {et~al.}(2023)Johnson, Gajjar, Keane, McKenna, Giese, McKeon,
  Carozzi, Alcaria, Brennan, Brzycki, Croft, Drew, Elkins, Gallagher, Kelly,
  Lebofsky, MacMahon, McCauley, Pater, Raeside, Siemion, \&
  Worden}]{johnson_simultaneous_2023}
Johnson, O.~A., Gajjar, V., Keane, E.~F., {et~al.} 2023, The Astronomical
  Journal, 166, 193, \dodoi{10.3847/1538-3881/acf9f5}

\bibitem[{{Kalberla} {et~al.}(2005){Kalberla}, {Burton}, {Hartmann}, {Arnal},
  {Bajaja}, {Morras}, \& {P{\"o}ppel}}]{LAB_survey_data}
{Kalberla}, P.~M.~W., {Burton}, W.~B., {Hartmann}, D., {et~al.} 2005, \aap,
  440, 775, \dodoi{10.1051/0004-6361:20041864}

\bibitem[{Kirsten {et~al.}(2022)Kirsten, Marcote, Nimmo, Hessels, Bhardwaj,
  Tendulkar, Keimpema, Yang, Snelders, Scholz, Pearlman, Law, Peters,
  Giroletti, Paragi, Bassa, Hewitt, Bach, Bezrukovs, Burgay, Buttaccio, Conway,
  Corongiu, Feiler, Forssén, Gawroński, Karuppusamy, Kharinov, Lindqvist,
  Maccaferri, Melnikov, Ould-Boukattine, Possenti, Surcis, Wang, Yuan,
  Aggarwal, Anna-Thomas, Bower, Blaauw, Burke-Spolaor, Cassanelli, Clarke,
  Fonseca, Gaensler, Gopinath, Kaspi, Kassim, Lazio, Leung, Li, Lin, Masui,
  Mckinven, Michilli, Mikhailov, Ng, Orbidans, Pen, Petroff, Rahman, Ransom,
  Shin, Smith, Stairs, \& Vlemmings}]{kirsten_repeating_2022}
Kirsten, F., Marcote, B., Nimmo, K., {et~al.} 2022, Nature, 602, 585,
  \dodoi{10.1038/s41586-021-04354-w}

\bibitem[{Kirsten {et~al.}(2024)Kirsten, Ould-Boukattine, Herrmann, Gawroński,
  Hessels, Lu, Snelders, Chawla, Yang, Blaauw, Nimmo, Puchalska, Wolak, \& van
  Ruiten}]{kirsten_link_2024}
Kirsten, F., Ould-Boukattine, O.~S., Herrmann, W., {et~al.} 2024, Nature
  Astronomy, 8, 337, \dodoi{10.1038/s41550-023-02153-z}

\bibitem[{{Lin} {et~al.}(2022){Lin}, {Lin}, {Li}, {Tseng}, {Jiang}, {Wang},
  {Cheng}, {Pen}, {Chen}, {Chen}, {Chen}, {Goto}, {Hashimoto}, {Hwang}, {King},
  {Kubo}, {Kuo}, {Mills}, {Nam}, {Oshiro}, {Shen}, {Tseng}, {Wang}, {Wu},
  {Bower}, {Chang}, {Chen}, {Chen}, {Chiang}, {Fedynitch}, {Gusinskaia}, {Ho},
  {Hsiao}, {Hu}, {Huang}, {J{\'a}uregui Garc{\'\i}a}, {Kim}, {Kuo}, {Ling},
  {On}, {Peterson}, {R. Raquel}, {Su}, {Uno}, {Wu}, {Yamasaki}, \&
  {Zhu}}]{BURSTT}
{Lin}, H.-H., {Lin}, K.-y., {Li}, C.-T., {et~al.} 2022, \pasp, 134, 094106,
  \dodoi{10.1088/1538-3873/ac8f71}

\bibitem[{{Manchester} {et~al.}(2005){Manchester}, {Hobbs}, {Teoh}, \&
  {Hobbs}}]{PsrCat}
{Manchester}, R.~N., {Hobbs}, G.~B., {Teoh}, A., \& {Hobbs}, M. 2005, \aj, 129,
  1993, \dodoi{10.1086/428488}

\bibitem[{McInnes {et~al.}(2017)McInnes, Healy, \& Astels}]{McInnes2017}
McInnes, L., Healy, J., \& Astels, S. 2017, Journal of Open Source Software, 2,
  205, \dodoi{10.21105/joss.00205}

\bibitem[{{Mereghetti} {et~al.}(2020){Mereghetti}, {Savchenko}, {Ferrigno},
  {G{\"o}tz}, {Rigoselli}, {Tiengo}, {Bazzano}, {Bozzo}, {Coleiro},
  {Courvoisier}, {Doyle}, {Goldwurm}, {Hanlon}, {Jourdain}, {von Kienlin},
  {Lutovinov}, {Martin-Carrillo}, {Molkov}, {Natalucci}, {Onori}, {Panessa},
  {Rodi}, {Rodriguez}, {S{\'a}nchez-Fern{\'a}ndez}, {Sunyaev}, \&
  {Ubertini}}]{integralFRB}
{Mereghetti}, S., {Savchenko}, V., {Ferrigno}, C., {et~al.} 2020, \apjl, 898,
  L29, \dodoi{10.3847/2041-8213/aba2cf}

\bibitem[{{Murphy} {et~al.}(2021){Murphy}, {Callanan}, {McCauley}, {McKenna},
  {Fionnag{\'a}in}, {Louis}, {Redman}, {Ca{\~n}izares}, {Carley}, {Maloney},
  {Coghlan}, {Daly}, {Scully}, {Dooley}, {Gajjar}, {Giese}, {Brennan}, {Keane},
  {Maguire}, {Quinn}, {Mooney}, {Ryan}, {Walsh}, {Jackman}, {Golden}, {Ray},
  {Doyle}, {Rigney}, {Burton}, \& {Gallagher}}]{ilofar}
{Murphy}, P.~C., {Callanan}, P., {McCauley}, J., {et~al.} 2021, \aap, 655, A16,
  \dodoi{10.1051/0004-6361/202140415}

\bibitem[{{Olausen} \& {Kaspi}(2014)}]{McGillMagnetars}
{Olausen}, S.~A., \& {Kaspi}, V.~M. 2014, \apjs, 212, 6,
  \dodoi{10.1088/0067-0049/212/1/6}

\bibitem[{{Petroff} {et~al.}(2019){Petroff}, {Hessels}, \&
  {Lorimer}}]{petroffreview}
{Petroff}, E., {Hessels}, J.~W.~T., \& {Lorimer}, D.~R. 2019, \aapr, 27, 4,
  \dodoi{10.1007/s00159-019-0116-6}

\bibitem[{{Rafiei-Ravandi} \&
  Smith(2023)}]{rafiei-ravandiMitigatingRadioFrequency2023}
{Rafiei-Ravandi}, M., \& Smith, K.~M. 2023, The Astrophysical Journal
  Supplement Series, 265, 62, \dodoi{10.3847/1538-4365/acc252}

\bibitem[{Ravi \& Collaboration(2023)}]{raviDSA110OverviewFirst2023}
Ravi, V., \& Collaboration, D.-. 2023, Bulletin of the AAS, 55

\bibitem[{{Rodriguez}(2025)}]{RodriguezLPRT}
{Rodriguez}, A.~C. 2025, \aap, 695, L8, \dodoi{10.1051/0004-6361/202553684}

\bibitem[{{Shah} {et~al.}(2025){Shah}, {Shin}, {Leung}, {Fong}, {Eftekhari},
  {Amiri}, {Andersen}, {Andrew}, {Bhardwaj}, {Brar}, {Cassanelli},
  {Chatterjee}, {Curtin}, {Dobbs}, {Dong}, {Dong}, {Fonseca}, {Gaensler},
  {Halpern}, {Hessels}, {Ibik}, {Jain}, {Joseph}, {Kaczmarek}, {Kahinga},
  {Kaspi}, {Kharel}, {Landecker}, {Lanman}, {Lazda}, {Main}, {Mas-Ribas},
  {Masui}, {Mckinven}, {Mena-Parra}, {Meyers}, {Michilli}, {Nimmo}, {Pandhi},
  {Patil}, {Pearlman}, {Pleunis}, {Prochaska}, {Rafiei-Ravandi}, {Sammons},
  {Sand}, {Scholz}, {Smith}, \& {Stairs}}]{chime_elliptical_2}
{Shah}, V., {Shin}, K., {Leung}, C., {et~al.} 2025, \apjl, 979, L21,
  \dodoi{10.3847/2041-8213/ad9ddc}

\bibitem[{{Sharma} {et~al.}(2024){Sharma}, {Ravi}, {Connor}, {Law}, {Ocker},
  {Sherman}, {Kosogorov}, {Faber}, {Hallinan}, {Harnach}, {Hellbourg}, {Hobbs},
  {Hodge}, {Hodges}, {Lamb}, {Rasmussen}, {Somalwar}, {Weinreb}, {Woody},
  {Leja}, {Anand}, {Das}, {Qin}, {Rose}, {Dong}, {Miller}, \& {Yao}}]{sharma24}
{Sharma}, K., {Ravi}, V., {Connor}, L., {et~al.} 2024, \nat, 635, 61,
  \dodoi{10.1038/s41586-024-08074-9}

\bibitem[{Tange(2025)}]{tange_2025_14911163}
Tange, O. 2025, GNU Parallel 20250222 ('Grete Tange'),  Zenodo,
  \dodoi{10.5281/zenodo.14911163}.
\newblock \url{https://doi.org/10.5281/zenodo.14911163}

\bibitem[{Thompson {et~al.}(2017)Thompson, Moran, \& Swenson}]{Thompson2017}
Thompson, A.~R., Moran, J.~M., \& Swenson, G.~W. 2017, Digital Signal
  Processing (Cham: Springer International Publishing), 309--390.
\newblock \url{https://doi.org/10.1007/978-3-319-44431-4_8}

\bibitem[{{van Haarlem} {et~al.}(2013){van Haarlem}, {Wise}, {Gunst}, {Heald},
  {McKean}, {Hessels}, {de Bruyn}, {Nijboer}, {Swinbank}, {Fallows},
  {Brentjens}, {Nelles}, {Beck}, {Falcke}, {Fender}, {H{\"o}randel},
  {Koopmans}, {Mann}, {Miley}, {R{\"o}ttgering}, {Stappers}, {Wijers},
  {Zaroubi}, {van den Akker}, {Alexov}, {Anderson}, {Anderson}, {van Ardenne},
  {Arts}, {Asgekar}, {Avruch}, {Batejat}, {B{\"a}hren}, {Bell}, {Bell}, {van
  Bemmel}, {Bennema}, {Bentum}, {Bernardi}, {Best}, {B{\^\i}rzan}, {Bonafede},
  {Boonstra}, {Braun}, {Bregman}, {Breitling}, {van de Brink}, {Broderick},
  {Broekema}, {Brouw}, {Br{\"u}ggen}, {Butcher}, {van Cappellen}, {Ciardi},
  {Coenen}, {Conway}, {Coolen}, {Corstanje}, {Damstra}, {Davies}, {Deller},
  {Dettmar}, {van Diepen}, {Dijkstra}, {Donker}, {Doorduin}, {Dromer}, {Drost},
  {van Duin}, {Eisl{\"o}ffel}, {van Enst}, {Ferrari}, {Frieswijk}, {Gankema},
  {Garrett}, {de Gasperin}, {Gerbers}, {de Geus}, {Grie{\ss}meier}, {Grit},
  {Gruppen}, {Hamaker}, {Hassall}, {Hoeft}, {Holties}, {Horneffer}, {van der
  Horst}, {van Houwelingen}, {Huijgen}, {Iacobelli}, {Intema}, {Jackson},
  {Jelic}, {de Jong}, {Juette}, {Kant}, {Karastergiou}, {Koers}, {Kollen},
  {Kondratiev}, {Kooistra}, {Koopman}, {Koster}, {Kuniyoshi}, {Kramer},
  {Kuper}, {Lambropoulos}, {Law}, {van Leeuwen}, {Lemaitre}, {Loose}, {Maat},
  {Macario}, {Markoff}, {Masters}, {McFadden}, {McKay-Bukowski}, {Meijering},
  {Meulman}, {Mevius}, {Middelberg}, {Millenaar}, {Miller-Jones}, {Mohan},
  {Mol}, {Morawietz}, {Morganti}, {Mulcahy}, {Mulder}, {Munk}, {Nieuwenhuis},
  {van Nieuwpoort}, {Noordam}, {Norden}, {Noutsos}, {Offringa}, {Olofsson},
  {Omar}, {Orr{\'u}}, {Overeem}, {Paas}, {Pandey-Pommier}, {Pandey}, {Pizzo},
  {Polatidis}, {Rafferty}, {Rawlings}, {Reich}, {de Reijer}, {Reitsma},
  {Renting}, {Riemers}, {Rol}, {Romein}, {Roosjen}, {Ruiter}, {Scaife}, {van
  der Schaaf}, {Scheers}, {Schellart}, {Schoenmakers}, {Schoonderbeek},
  {Serylak}, {Shulevski}, {Sluman}, {Smirnov}, {Sobey}, {Spreeuw}, {Steinmetz},
  {Sterks}, {Stiepel}, {Stuurwold}, {Tagger}, {Tang}, {Tasse}, {Thomas},
  {Thoudam}, {Toribio}, {van der Tol}, {Usov}, {van Veelen}, {van der Veen},
  {ter Veen}, {Verbiest}, {Vermeulen}, {Vermaas}, {Vocks}, {Vogt}, {de Vos},
  {van der Wal}, {van Weeren}, {Weggemans}, {Weltevrede}, {White}, {Wijnholds},
  {Wilhelmsson}, {Wucknitz}, {Yatawatta}, {Zarka}, \& {Zensus}}]{lofar}
{van Haarlem}, M.~P., {Wise}, M.~W., {Gunst}, A.~W., {et~al.} 2013, \aap, 556,
  A2, \dodoi{10.1051/0004-6361/201220873}

\bibitem[{{van Straten} {et~al.}(2021){van Straten}, Jameson, \&
  Osłowski}]{vanstratenPSRDADADistributedAcquisition2021}
{van Straten}, W., Jameson, A., \& Osłowski, S. 2021, Astrophysics Source Code
  Library, ascl:2110.003

\bibitem[{Vine \& Saji(2004)}]{SkyTemp-Vine}
Vine, D., \& Saji, A. 2004, Geoscience and Remote Sensing, IEEE Transactions
  on, 42, 119 , \dodoi{10.1109/TGRS.2003.817977}

\bibitem[{Weinreb \& Shi(2021)}]{weinrebLowNoiseAmplifier2021a}
Weinreb, S., \& Shi, J. 2021, IEEE Transactions on Microwave Theory and
  Techniques, 69, 2345, \dodoi{10.1109/TMTT.2021.3061459}

\end{thebibliography}
\bibliographystyle{aasjournal}

\appendix

\section{Telescope Calibration}\label{sec:Acalibration}
The factors of importance for determining the instrument sensitivity are three-fold: (1) the telescope response to an incoming signal as a function of angle, (2) the gain of the receiver, and (3) the system temperature of the instrument. 
Noise is added to an incoming signal during its propagation through the amplifiers and other electronics of the system. 
It is practical to begin by considering the specific intensity ($I_\nu$) of a signal, since it is a conserved quantity through empty space. 
However, the quantity that describes the spectral power of the source received at the detector is the flux density,
\begin{equation}
    S_\nu=\int_\Omega d\Omega\left[I_\nu(\theta,\phi)B(\theta)\cos\theta\right].\label{eq:flux_density}
\end{equation}
$B(\theta)$ describes the response of the receiver to incoming radiation and is equivalent to the normalized effective area $A_{e}(\theta)/A_0$, where $A_0=A_e(\theta=0)=A_{e,max}$. 
In this notation, $\theta=0$ points along the normal vector of the telescope receiver and $\phi$ describes the azimuthal angle about that normal vector.
The $\cos\theta$ term in the integral describes the effect of projecting the incoming spectral power across the detector at different inclination angles. 
The beam size of a telescope is relatively small in most cases, so $B(\theta)$ falls to zero rapidly and $\cos\theta\approx1$, but this approximation does not hold for the GReX instrument. 
Since the behavior of the $\cos\theta$ term describes the Flux Density seen at the detector, it is absorbed into the beam response function $B(\theta)$.
It is standard practice to express the specific intensity ($I_\nu$) of astrophysical signals in terms of their equivalent brightness temperature ($T_b$) using the Raleigh-Jeans approximation:
\begin{equation}
    I_\nu \approx \frac{2k_B T_b}{\lambda^2}. \label{eq:RJ}
\end{equation}
For a given distribution of brightness temperature on the sky ($T_b(\theta,\phi)$), the detected flux density is 
\begin{equation}
    S_\nu=\frac{2k_B}{\lambda^2}\int_\Omega d\Omega\left[T_b(\theta,\phi)B(\theta)\right].\label{eq:FDT}
\end{equation}
We can prescribe a constant effective temperature ($T_{eff}$) that gives the same flux density as for an arbitrary brightness distribution:
\begin{equation*}
    \int_\Omega d\Omega\left[T_b(\theta,\phi)B(\theta)\right]=T_{eff}\int_\Omega d\Omega\left[B(\theta)\right]=T_{eff}\Omega_B.
\end{equation*}
For a single-pixel instrument like GReX, this is equivalent to the antenna temperature contributed by the source. 
If the source does have a constant brightness temperature across the beam, then it is apparent that $T_{eff}$ accurately describes the source temperature. 
However, for sources covering a small patch of sky ($T_b(\theta,\phi)=T_b\Delta^\prime$, where$\Delta^\prime\sim\delta(\theta-\theta^\prime, \phi-\phi^\prime)$) centered at $(\theta^\prime,\phi^\prime)$ and with total solid angle $\Omega_\Delta$, the beam response and source brightness temperature are approximately constant and
\begin{equation*}
    T_{eff}\approx\frac{\int_\Omega T_b(\theta,\phi)B(\theta)\Delta^\prime d\Omega}{\Omega_B} =T_b B(\theta^\prime)\frac{\Omega_\Delta}{\Omega_B}.
\end{equation*}
Since the flux density is related to the effective temperature like
\begin{equation*}
    S_\nu=\frac{2k_B\Omega_B}{\lambda^2}T_{eff},
\end{equation*}
a source at a distance $\delta\theta$ from the center of the beam will have its flux density at the receiver reduced by a factor $B(\delta\theta)$ than if it were centered in the beam:
\begin{equation*}
    S_\nu(\delta\theta) \approx B(\delta\theta)S_\nu(0). \label{eq:FnuAstro}
\end{equation*}
While flux density is usually not used to describe extended sources with variable brightness temperature, we do use it for calibration of the instrument as described in \autoref{sec:beam_model} with measured and simulated values of the hydrogen I line brightness temperature from diffuse Galactic gas. 
As such, we include an expression for the antenna temperature for a well-known brightness distribution:
\begin{equation}
    T_{A}=\int_\Omega d\Omega \left[T_b(\theta,\phi)\hat{B}(\theta)\right],\label{eq:Teff}
\end{equation}
where $\hat{B}$ describes the normalized telescope response,
\begin{equation}
    \hat{B}(\theta)=\frac{B(\theta)}{\Omega_B}.\label{eq:Bnorm}
\end{equation}
The effective area of the telescope has the intrinsic property that
\begin{equation*}
\langle A_e\rangle_\Omega=\frac{\int_\Omega}{4\pi}=\frac{\lambda^2}{4\pi}.
\end{equation*}
Following from this, and $B(\theta)=A_e(\theta)/A_0$,
the maximum effective area is 
\begin{equation}
    A_0=\frac{\lambda^2}{\Omega_B}.\label{eq:Ae}
\end{equation}
The forward gain ($g_f$) relates the antenna temperature to the source flux density as
\begin{equation*}
S_\nu=g_f T_{eff} \to g_f=\frac{2k_B\Omega}{\lambda^2},\label{eq:FDFG}
\end{equation*}
equivalent with
\begin{equation}
    g_f=\frac{2k_B}{A_0}\label{eq:FG}
\end{equation}
Since the forward gain relates the observed temperature of a source by the instrument to the flux density of that source, it is a critical component for defining the sensitivity of the instrument. 

Due to the prevalence of describing observed signals in terms of $T_{A}$, it is convenient to describe all sources that contribute additive power to the signal within the system as a system temperature ($T_{sys}$), even though the noise is not necessarily thermal in nature. 
While there are many sources that contribute to $T_{sys}$, we focus on the thermal noise within the receiver electronics ($T_{rec}$) and the background temperature of the sky ($T_{sky}$). 
At any time, the total observed temperature by the system is
\begin{equation}
    T_{obs}=T_{eff}+T_{sys}=T_{eff}+T_{rec}+T_{sky}.
\end{equation}
However, GReX data is natively stored as linear intensities with arbitrary units ($I_{arb}$). 
The receiver gain ($G$, in $K/arb$) converts the linear intensities into Kelvin,
\begin{equation}
    T_{obs}=GI_{arb}.
\end{equation}
Since $T_{sky}$ is well characterized to account for atmospheric effects and background radiation, our sensitivity analysis needs to account for $G$ and $T_{sys}$ so that the effective source temperature can be isolated as
\begin{equation}
    T_{eff}=GI_{arb}-T_{sky}-T_{rec}.
\end{equation}
We determine $G$ and $T_{sys}$ by performing a Y-factor test with two known values for $T_{sky}$.
We first take measurements of the unobscured sky with a GReX terminal before collecting data again, this time with a slab of room-temperature radio-absorbent foam covering the receiver. 
Under the assumption that there are no significant astrophysical sources present in the data, we now have 
\begin{equation}
    I_1=\frac{T_{sky}+T_{rec}}{G},\\
    I_2=\frac{T_{hot}+T_{rec}}{G}.
\end{equation}
The Y-factor ($Y=I_2/I_1$) is then expanded according to the above equations and rearranged to give the following expression for the receiver temperature,
\begin{equation}
    T_{rec}=\frac{T_{hot}-YT_{sky}}{Y-1}.\label{eq:Trec}
\end{equation}
From this value, the receiver gain is trivially
\begin{equation}
    G=\frac{T_{hot}+T_{rec}}{I_2}=\frac{(T_{hot}-T_{sky})}{(I_2-I_1)}.\label{eq:gain}
\end{equation}
Ideally, this receiver gain is roughly constant over observing epochs, but changes in temperature and degradation of the electronics can cause slow changes in this value. 
Of course, any changes to the programmable gain in the FEM of a GReX terminal will require remeasuring the receiver gain. 
The more volatile value is the system temperature, which can change depending on environmental effect such as prevalence of RFI and accumulation of water in the feed, among other potential issues. 
As such, it is useful to measure $T_{sys}$ semi-regularly. 

Once $G$ and $T_{sys}$ are well characterized for the terminal, we compare the measured $T_{eff}$ of a source to its known flux density to determine the forward gain. 
We utilize all-sky maps of neutral galactic hydrogen to simulate the expected flux density as seen by the GReX terminal and compare this with the measured $T_{eff}$ of the HI line by the instrument to calibrate $g_f$ in \autoref{sec:beam_model}.
The conversion of temperature units into flux densities is commonly applied to the system temperature of an instrument to define its system equivalent flux density (SEFD),
\begin{equation}
    \text{SEFD}=S_{\nu,sys}=g_fT_{sys}.\label{eq:sefd}
\end{equation}
Since we are dealing with dynamic spectra that contain signals across multiple frequency channels, time samples, and polarizations, it is helpful to consider the power and temperature within the detector noise that limits the detectability of a signal. 
A signal spanning a bandwidth $\Delta\nu$ with total integration time $\tau$ that is seen in both polarizations of a dipole antenna will be present in $N=2\Delta\nu\tau$ total samples. 
The detectability of such a signal depends on the signal-to-noise ratio
\begin{equation}
    \text{SNR}=\frac{T_{src}}{T_{rms}},
\end{equation}
where $T_{rms}=\sigma_T/\sqrt{N}$ describes the root-mean-squared noise in the system temperature. 
Assuming that the thermal noise contributing to $T_{sys}$ is exponentially distributed such that $T_{sys}=\mu_T=\sigma_T$, then 
\begin{equation*}
T_{rms}=\frac{T_{sys}}{\sqrt{N}}.
\end{equation*}
We can then rewrite the signal-to-noise ratio as
\begin{equation*}
    \text{SNR}=\frac{T_{src}}{T_{sys}}\sqrt{N}
\end{equation*}
The SNR is a useful measure since this ratio of temperatures is equivalent to the corresponding ratios of native system intensity units and the ratios of Flux Density. 
Converting the above expression to flux densities gives
\begin{equation*}
    S_{\nu,src}=\text{SNR}\frac{\text{SEFD}}{\sqrt{N}}.
\end{equation*}
The flux density threshold for detection is then given by
\begin{equation*}
    S_{\nu,min}=\text{SNR}_{min}\cdot\text{NEFD},
\end{equation*}
where the NEFD (noise-equivalent flux density) is $\text{NEFD}=g_fT_{rms}=\text{SEFD}/{\sqrt{N}}$.

\section{HI Line Simulation}\label{sec:Asimulation}
The LAB data is an all-sky map of the brightness temperatures of neutral galactic hydrogen binned into discrete velocity channels. 
These scalar velocities represent the component of the motion of the neutral galactic hydrogen along the line of sight from the solar system to the gas within the local standard of rest (LSR) frame.
The original data format for the LAB survey had a velocity channel spacing of $\sim 1.031\text{ km/s}$ and a total velocity range of $-450\text{ km/s}$ to $+450\text{ km/s}$, but this data was conglomerated into wider bins with a spacing of $10\text{ km/s}$. 
This was done by averaging the brightness temperatures of the narrower bins into those respective larger bins.
Thus, the LAB all-sky map is stored in \textit{HEALPix} \citep{HEALPix} file format with separate files for the difference wide velocity channels. 
Each \textit{HEALPix} velocity-channel file has data stored in two fields: the first is a \textit{TEMPERATURE} field, which gives the brightness temperature within the channel for each sky position pixel; the second is called the \textit{SIMULATION} field, which gives the number of original $\sim 1.031\text{ km/s}$ velocity channels that were used in calculating that brightness temperature. 
From the \textit{TEMPERATURE} field, we construct a map of brightness temperature, $T_{HI}(\ell_j, b_j,v_k)$, where $(\ell_j, b_j)$ describes the central position of the $j^{th}$ pixel in galactic coordinates and $v_k$ gives the $k^{th}$ velocity channel in the LSR frame. 
To construct an expected spectrum of HI from this map for a specific terminal and observing time, we need to: (1) account for the velocity, $\vec{v}_{obs}$, and central pointing, $(\ell,b)_{obs}$, of the terminal in the LSR frame at that time; (2) remove the velocity of the observer in LSR along the LoS to each pixel from the neutral hydrogen velocity in LSR to get the gas velocity in the terminal's frame; (3) compute the angular deviation ($\delta\theta)$ of each pixel from the central pointing of the terminal feed and use that to construct a simulated 2D Gaussian beam response $B(\delta\theta)$; (4) convert hydrogen velocity in the observer frame to HI line frequency ($f_{HI,0}=1420.40575\text{ MHz}$) and bin frequencies into observing channels; and (5) for each frequency channel, integrate the brightness temperature map over the $4\pi$ steradians of sky modulated by the telescope beam response function $(B(\delta\theta,f))$ to get the expected spectrum of brightness temperature of HI as seen from a GReX unit.

To compute the velocity of neutral hydrogen in the reference frame of the GReX terminal along the LoS to each LAB data pixel, we need to start with an expression for the overall gas velocity in the observer frame:
\begin{equation*}
    \vec{v}_H,obs(t) = \vec{v}_{H, LSR} - \vec{v}_{obs, LSR}(t).
\end{equation*}
Then, the velocity of gas along the LoS to the $j^{th}$ data pixel is
\begin{equation*}
    v_{H,obs,j}(t) = \hat{n}_j \cdot (\vec{v}_{H, LSR} - \vec{v}_{obs, LSR}(t)),
\end{equation*}
where $\hat{n}_j$ is the unit vector pointing along that LoS.
The scalar velocity channels of the LAB data are already the gas velocity along the LoS, so our final expression is
\begin{equation*}
    v_{H,j,k,obs}(t) = v_{H,k,LSR} - v_{obs,j,LSR}(t),
\end{equation*}
where $v_{obs,j,LSR}(t)=\hat{n}_j\cdot\vec{v}_{obs, LSR}(t)$ is the component of observer velocity along the $j^{th}$ LoS.
\textit{Astropy} handles the conversion of sky positions and velocities between different reference frames and coordinate systems to generate values for $v_{obs,j,LSR}$. 
We begin by defining an \textit{EarthLocation} object for the terminal in geodetic coordinates ($x_{obs,geo}(\text{lat},\text{lon})$), converting to a location in the barycentric celestial reference system at a specific time ($t$), which is then transformed directly into ICRS and then LSR coordinates ($x_{obs,LSR}(x_{obs,geo},t)$). 
\textit{Astropy} then computes the Cartesian differential of the \textit{EarthLocation} object to determine the the GReX terminal velocity within the LSR frame ($\vec{v}_{obs,LSR}(t)$) in (x,y,z) coordinates. 
The LoS unit vector $\hat{n}_j=\hat{n}(\ell_j,b_j)$ that points from the observer towards the galactic coordinates $\ell_j$ and $b_j$ is generated by converting an \textit{Astropy} \textit{SkyCoord} object at those galactic coordinates into (x,y,z) to remain consistent with $x_{obs,LSR}$ and $\vec{v}_{obs,LSR}(t)$. 
The terminal's velocity within the LSR frame along each LoS towards an individual LAB data pixel is then computed as the dot product
\begin{equation*}
    v_{obs,j,LSR}(t) =  \hat{n}_j\cdot\vec{v}_{obs,LSR}(t).
\end{equation*}
The frequency of the HI line as seen by the GReX terminal along the $j^{th}$ LoS and from the $k^{th}$ LAB data velocity channel are computed from this scalar gas velocity according to
\begin{equation*}
    f_{HI,j,k} = \gamma_{j,k}(1-\beta_{j,k})f_{HI,0},
\end{equation*}
where $\beta_{j,k}=v_{obs,j,LSR}/c$, $\gamma_{j,k}=1/\sqrt{1-\beta_{j,k}^2}$, where $f_{HI,0}$ is the lab-frame frequency of emission for neutral hydrogen gas. 
After applying these transformations, our map of brightness temperature is formatted as $T_{HI}(\ell_j,b_j,f_{j,k})$. 
To standardize the data, we consider a hypothetical observing frequency channelization scheme labeled as $f_i$ such that channel edges directly abut one another. 
The temperature map is then reconfigured such that
\begin{equation*}
    T_{HI}(\ell_j, b_j, f_i, t) = \sum_{k}T_{HI}(\ell_j,b_j,f_{j,k}(t))|f_{j,k}(t)\in f_i|,
\end{equation*}
and we now have an all-sky map of brightness temperature within each frequency channel at each observing epoch.

The actual brightness temperature seen by a GReX terminal is found by integrating the brightness temperature of your source over the normalized beam response function of the feed:
\begin{equation*}
    \widetilde{T}_{HI}(t,f) = \frac{\int_\Omega d\Omega \left[T_{HI}(\ell, b, f, t) B(\ell(t),b(t))\right]}{\int_\Omega d\Omega B(\ell(t),b(t))}.
\end{equation*}
In our case, we replace this integral with its numerical counterpart,
\begin{equation*}
    \widetilde{T}_{HI}(t,f_i) \approx \sum_j \Delta\Omega\left[T_{HI}(\ell_j, b_j, f_i, t) \hat{B}(\ell_j(t),b_j(t))\right],
\end{equation*}
which requires that we compute the normalized beam response function for a GReX terminal at each LAB pixel galactic coordinate as a function of time. 
We model the normalized beam response as a two-dimensional Gaussian,
\begin{equation*}
    \hat{B}(\theta,\theta_\text{FWHM})=\Omega_{B}^{-1}e^{-4\ln2[\theta/\theta_\text{FWHM}]^2},
\end{equation*}
with a full-width at half maximum of $\theta_\text{FWHM}$. 
It is important to note that the $\Omega_{beam}\propto f^{-2}$ relationship according to the radiometer equation means that the beam needs to be simulated for every frequency channel. 
We numerically integrate over simulated beams across a range of different $\theta_\text{FWHM}$and interpolate between the resulting $\Omega_B(\theta_\text{FWHM})$. 
We consider a characteristic beam width $\Omega_B(f_{HI,0})$, which can be scaled across the band according to the inverse squared frequency dependence. 
We assume that $\Omega_B\approx\Omega_B(f_{HI,0})$ in the $\sim2\text{MHz}$ of band used in this analysis. 
We parametrize the angle dependence of the beam response function into galactic coordinates such that $\theta\to\delta\theta(\ell_j,b_j)=\delta\theta_j$, where $\delta\theta$ describes the off-axial deviation of the $j^{th}$ data pixel coordinate from the beam central pointing $(\ell,b)_{obs}$. 
Thus,
\begin{equation*}
    \hat{B}(\theta,\theta_\text{FWHM})\to \hat{B}(\delta\theta_j, \theta_\text{FWHM}(f))=\Omega_{B}^{-1}e^{-4\ln2[\delta\theta_j/\theta_\text{FWHM}(f)]^2}
\end{equation*}
The normalizing factor $\Omega_B$ is found by numerically integrating the beam response over the full $4\pi$ steradians of sky at the \textit{HEALPix} data coordinates,
\begin{equation*}
    \Omega_{B}(\theta_\text{FWHM}(f)) \approx \left[\Delta\Omega\sum_j e^{-4\ln2\left(\frac{\delta\theta_j}{\theta_\text{FWHM}(f)}\right)^2}\right].
\end{equation*}
The central pointing of the GReX feed is needed to calculate $\delta\theta_j$.
We first define an \textit{AltAz} frame for the terminal at the appropriate \textit{EarthLocation} and time, $t$ which is then fed into a \textit{SkyCoord} object with the corresponding altitude and azimuth values for the feed (which should be $90^\circ$ and $0^\circ$, respectively). 
We then extract the galactic coordinates that correspond to this \textit{SkyCoord} object, which acts as the central pointing of the feed, $(\ell,b)_{obs}(t)$. 
The difference in angle ($\delta\theta$) between the $j^{th}$ data point and the terminal pointing in galactic coordinates is calculated according to 
\begin{equation*}
    \cos{\delta\theta_j(t)} = \cos{[90^\circ-b_{obs}(t)]}\cos{[90^\circ-b_j]} + \sin{[90^\circ-b_{obs}(t)]}\sin{[90^\circ-b_j]}\cos{[\ell_j-\ell_{obs}(t)]}.
\end{equation*}
We include a small-angle approximation for numerical stability, which gives the resulting angular distance as 
\begin{equation*}
    \delta\theta_{j}(t) = 
    \begin{cases}
        \sqrt{2[1-\cos\delta\theta_j(t)]},& |1-|\cos\delta\theta_j(t)||<10^{-6}\\
        \arccos{[\cos\delta\theta_j(t)]},& \text{else}
    \end{cases}.
\end{equation*}
Using this definition for $\delta\theta_j(t)$, we generate an all-sky map of the beam response at each \textit{HEALPix} data coordinate for all observing times and frequency channels,
\begin{equation*}
    B_{\theta_{\text{FWMH}, c},f_c}(\ell_j,b_j,f_i,t)=\exp{\left[-4\ln2\left(\frac{\delta\theta_j(t)}{\theta_\text{FWHM}(f_i)}\right)^2\right]},
\end{equation*}
and compute the normalizing factor as $\Omega_B(t,f_i)=\Delta\Omega\sum_jB_{\theta_{\text{FWHM},c},f_c}(\ell_j,b_j,f_i,t)$.
Substituting the expressions for $\theta_\text{FWHM}(f)$ and $\delta\theta_j(t)$ into the earlier expressions for $B$ and $\Omega_B$ yields the normalized beam response ($\hat{B}$) as a sky map at the \textit{HEALPix} galactic coordinates for each observing time and frequency channel.

We can now apply this set of simulated beam responses to the all-sky temperature map ($T_{HI}(\ell_j,b_j,f_i,t)$) and sum over the $j$ data coordinates to generate the expected observed spectrum of neutral galactic hydrogen in temperature units as seen from the GReX terminal at each observing time:
\begin{equation*}
    \widetilde{T}(t,f_i) = \frac{\Delta\Omega}{\Omega_B(t,f_i)}\sum_j \left[T_{HI}(\ell_j,b_j,f_i,t)B_{\theta_{\text{FWHM},c},f_c}(\ell_j,b_j,f_i,t)\right].
\end{equation*}
In \autoref{sec:beam_model}, we use the method described above to generate the expected spectra of the HI-line for terminals at OVRO and Cornell University. 
We find that integrating across the spectrum at each observation time and comparing the total HI-line intensity was sufficient for determining the size of the terminal beams. 

\end{document}